%% file: main.tex
\documentclass[notitlepage,12pt]{jedm}
\usepackage[table]{xcolor}
\usepackage{amsfonts,amsmath,amssymb,amsthm}
\usepackage{url}
\usepackage{hyperref}
\hypersetup{
  colorlinks   = true, 
  urlcolor     = blue, 
  linkcolor    = blue, 
  citecolor   = blue 
}
\usepackage{graphicx}

\usepackage{array}
\usepackage{multirow}

\newcolumntype{L}[1]{>{\raggedright\let\newline\\\arraybackslash\hspace{0pt}}m{#1}}
\newcolumntype{C}[1]{>{\centering\let\newline\\\arraybackslash\hspace{0pt}}m{#1}}
\newcolumntype{R}[1]{>{\raggedleft\let\newline\\\arraybackslash\hspace{0pt}}m{#1}}

\usepackage{bm}

\newcommand{\EE}{\mathbb{E}}
\newcommand{\VV}{\mathbb{V}}

\newcommand{\rct}{RCT}
\newcommand{\rem}{REM}
\newcommand{\pop}{POP}

\newcommand{\pate}[1][\pop]{\EE_{#1}[\tau]}
\newcommand{\sate}[1][\rct]{\overline{\tau}_{#1}}
\newcommand{\patesub}[2][\pop]{\EE_{#1}[\tau\mid #2]}

\newcommand{\tdm}{\hat{\tau}^{DM}}

\newcommand{\alg}{\predr[(0)]{\cdot;\bm{\beta}}}

\newcommand{\rebar}{\hat{\tau}_{rebar}}
\newcommand{\myloop}{\hat{\tau}_{LOOP}(\bm{x})}

\newcommand{\yc}{y(0)}
\newcommand{\yt}{y(1)}
\newcommand{\yci}{y_i(0)}
\newcommand{\yti}{y_i(1)}

\newcommand{\pred}[2]{\widehat{y(#1)}^{\rct}\left(#2\right)}
\newcommand{\predic}{\widehat{\yci}^{\rct}}
\newcommand{\predit}{\widehat{\yti}^{\rct}}

\newcommand{\predr}[2][]{\hat{y}^{\rem}#1\left(#2\right)}
\newcommand{\predri}[1][i]{\hat{y}_{#1}^r}
\newcommand{\reloop}{\hat{\tau}_{LOOP}(\bm{\predri[]},\bm{x})}
\newcommand{\reloopV}{\hat{\tau}_{LOOP}(\bm{\predri[]})}

\begin{document}

\title{Using Auxiliary Data to Boost Precision in the Analysis of A/B Tests on an Online Educational Platform: New Data and New Results\footnote{Data and code used in this work can be found at \url{https://osf.io/k8ph9/}.}}
\date{} 

\author{
{\large Adam C. Sales}\\Worcester Polytechnic Institute\\asales@wpi.edu\\ \and {\large Ethan B. Prihar}\\Worcester Polytechnic Institute\\ebprihar@gmail.com\\ \and {\large Johann A. Gagnon-Bartsch}\\University of Michigan\\johanngb@umich.edu\\ 
\and \hskip 5em {\large Neil T. Heffernan}\\\hskip 5em Worcester Polytechnic Institute\\\hskip 5em nth@wpi.edu\\
}

\maketitle

\begin{abstract}
Randomized A/B tests within online learning platforms represent an exciting direction in learning sciences. With minimal assumptions, they allow causal effect estimation without confounding bias and exact statistical inference even in small samples. However, often experimental samples and/or treatment effects are small, A/B tests are underpowered, and effect estimates are overly imprecise. Recent methodological advances have shown that power and statistical precision can be substantially boosted by coupling design-based causal estimation to machine-learning models of rich log data from historical users who were not in the experiment. Estimates using these techniques remain unbiased and inference remains exact without any additional assumptions. This paper reviews those methods and applies them to a new dataset including over 250 randomized A/B comparisons conducted within ASSISTments, an online learning platform. We compare results across experiments using four novel deep-learning models of auxiliary data and show that incorporating auxiliary data into causal estimates is roughly equivalent to increasing the sample size by 20\% on average, or as much as 50-80\% in some cases, relative to t-tests, and by about 10\% on average, or as much as 30-50\%, compared to cutting-edge machine learning unbiased estimates that use only data from the experiments. We show that the gains can be even larger for estimating subgroup effects, hold even when the remnant is unrepresentative of the A/B test sample, and extend to post-stratification population effects estimators. \\

{\parindent0pt
\textbf{Keywords:} A/B tests, deep learning, evaluation
}
\end{abstract}

\section{Introduction}
In randomized A/B tests on an online learning platform, students are randomized between different educational conditions or strategies, and their subsequent educational outcomes of interest are compared between different conditions.
For instance, \citeN{harrison2020spacing} studied data from 2,152 middle- and high-school students whose teachers assigned a specific module---a ``skill builder''---on the ASSISTments online tutoring platform \cite{heffernan2014assistments}. Prior to the students' work, the authors designed four different educational conditions, which differed in how the numbers and symbols in arithmetic expressions were spaced. As students logged on to the platform, during their usual schoolwork, they were each individually randomized to one of the four conditions, and completed their work under that condition. Subsequently, the authors of the study compared the average number of problems students in each condition had to work on before achieving mastery, defined as answering three problems correctly in a row. They found that students who were assigned the ``congruent'' condition---in which the spacing between numbers corresponded to the order of operations---needed to work on roughly one fewer problem, on average, than students in the ``incongruent'' condition. This finding, and others reported in the paper, validated their previous scientific hypotheses regarding embodied cognition, the relationship between abstract learning, and the arrangement of objects in physical (or virtual) space.

In general, A/B tests have two significant advantages over observational study designs, which do not include randomization, and additional advantages over studies conducted in a lab. First, they are (famously) free of confounding bias---since students are randomly allocated between conditions, differences in outcomes must be due to either a causal effect of the randomized conditions or to random error, but not to systematic baseline differences between students, observed or unobserved. Perhaps less famously, randomization forms a ``reasoned basis for inference'' \cite{fisher1935design}: the (known) probabilities of allocation of students between experimental conditions provide nearly all of the necessary justification for the unbiased estimation of causal effects, as well as standard errors, confidence intervals, and p-values. No other distributional assumptions or modeling assumptions are necessary. These properties allowed \citeN{harrison2020spacing} to estimate causal effects of spacing conditions, as well as to statistically rule out other alternative explanations.\footnote{Actually the authors of that paper did make modeling assumptions in their analysis, but they could have conducted a non-parametric analysis.} 
Causal effect and standard error estimators that rely only on the experimental design are referred to as ``design-based'' \cite{schochet2015statistical}. 

On the other hand, A/B tests can be hobbled by statistical imprecision. For instance, \citeN{harrison2020spacing} was unable to confirm or disconfirm one of their initial hypotheses, regarding differences in causal effects between subgroups of students, because the standard errors of the relevant estimates were too high. Unlike observational studies using data from online tutors, the sample size in A/B tests is necessarily limited to those students who worked on the relevant modules while the study was taking place. In contrast, a typical observational study would use data from all students who have ever worked on the relevant modules, including the (often large) number of students who worked on them before the onset of the study, and might sometimes use data from students who worked on similar modules as well. Analysis of A/B tests must discard data from these students, who were not randomized between treatment conditions and are subject to confounding. 
Unlike studies conducted in carefully controlled laboratory environments, A/B tests are subject to the haphazard unpredictability of real life, which only increases statistical imprecision---even a sample as large as the 2,152 of \citeN{harrison2020spacing} may not be enough to answer some causal questions. 

However, recent methodological innovations \cite{gagnon2021precise,rebarEDM} have argued that data from the ``remnant'' from an experiment---students who were not randomized between conditions, but for whom covariate and outcome data are available---need not be discarded, but can play a valuable role in causal estimation. In fact, researchers can use data from the remnant to decrease experimental standard errors without sacrificing the unbiased estimation and design-based inference that recommend A/B testing. The basic idea is to first use the remnant data to train a machine learning model predicting outcomes as a function of covariates; then, use that fitted model to generate predicted outcomes for participants in the experiment. Finally, use those predictions as a covariate in a design-based covariate-adjusted causal estimator \cite[for eg.]{aronowMiddleton,wager,loop,chernozhukov2018double}. Variants of the method use the predictions from the remnant alongside other covariates to estimate causal effects.

 These methods can help alleviate another weakness, shared by A/B tests and observational studies---the dependence of conclusions on statistical modeling choices. By observing outcome data prior to selecting and fitting statistical models, researchers (often inadvertently) choose models most favorable to their desired conclusions and undermine statistical objectivity and the logic of inference. Two proposed solutions to this issue are (1) to split the sample prior to data analysis and use one part to choose a model and the second part to estimate effects \cite{heller2009split} or (2) to rely on flexible non-parametric models that can be specified prior to data collection \cite{van2011targeted}. Design-based estimators incorporating remnant data rely on both these techniques: model-fitting in the remnant can be interactive and based on human judgment, without adversely affecting the objectivity or validity of statistical inference using the experimental sample. Design-based covariate adjustment often uses robust or non-parametric models. 

This paper reviews design-based effect estimation from A/B tests, along with a set of design-based causal estimators that use remnant data (Section \ref{sec:background}). 
Next (Section \ref{sec:data}) we describe a new dataset that we used to test these methods: a collection of 68 multi-armed A/B tests run on the ASSISTments TestBed \cite[now called ``E-Trials'']{ostrow2016assessment}, which together include 227 different two-way comparisons and 38,035 students. Alongside this experimental data, we collected log data for an additional 193,218 students who worked on similar skill builders in ASSISTments but did not participate in any of the 68 experiments---the remnant.
The following section (Section \ref{sec:imputation}) describes the Deep Learning model that we trained in the remnant to predict student outcomes as a function of prior log data. 

The next four sections use that data and those models to address four research questions regarding the use of remnant data to assist in the analysis of A/B tests. 
The first research question (Section \ref{sec:rq1}) regards the overall efficacy of our approach: to what extent might remnant data improve the precision of effect estimates from A/B tests? Does it ever harm precision, in practice?
As part of this research question, we also investigated the roles that various types of remnant data may play in the process.
The second research question (Section \ref{sec:subgroup}) regards subgroup effects---treatment effects may be present for some groups of students but not others, or may differ between groups of students. However, breaking A/B test data into subsets further exacerbates sample size issues---is this something remnant data may help with? 
The third research question (Section \ref{sec:rq3}) regards differences between the remnant and A/B testing data---in particular, what if the remnant is known to be drawn from a different population than the participants in A/B tests? Can it still be useful? To answer this question, we purposely constructed a new remnant that we believe is composed mostly of white and Asian males and used it to analyze A/B testing data from primarily other demographic groups.
The last research question (Section \ref{sec:rq4}) asks if remnant data may be helpful in generalizing effects estimated from an A/B test to a wider population, even when subjects in the A/B test were not randomly drawn from that population.

Across the board, we find that estimates using the remnant are often substantially more precise than estimates that do not, and very rarely are much less precise. This holds for overall estimates, estimated subgroup effects, population average effects, and even when the remnant is unrepresentative of the A/B test by construction. 
Our results give a much clearer picture of the potential impacts of using remnant data in design-based causal inference than was previously available. 


\section{Background}\label{sec:background}
\subsection{Framework: Different (Groups of) Users, Different (Average) Treatment Effects}

\begin{figure}
    \centering
    \includegraphics{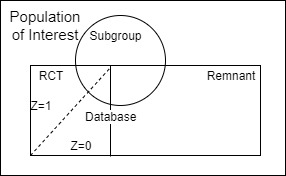}
    \caption{A Venn Diagram for the sets of subjects described in the text and Table \ref{tab:sets}.}
    \label{fig:sets}
\end{figure}

For the method we are describing, it will be useful to define several different sets of subjects or users, summarized in Table \ref{tab:sets} and Figure \ref{fig:sets} (also see \cite{imbens2004nonparametric}).

\begin{table}
\centering
\caption{Descriptions of sets of subjects described in the text, and associated causal estimands.}\vspace*{1ex}
\begin{tabular}{|c|c|p{.3\linewidth}|c|}
\hline
Name&Abbreviation& Explanation &Avg. Effect\\
\hline
RCT Set & $\rct$ & Participants in the RCT, randomized between $Z=0$ and $Z=1$ conditions &$\sate$\\
Population of interest & $\pop$ & The total population for which researchers wish to estimate effects &$\pate$\\
Subgroup $k$ & $G=k$ & One of $K$ disjoint subsets of $\rct$ or $\pop$ & \begin{tabular}{@{}c@{}} $\sate[G=k]$ or\\ $\patesub{G=k}$\end{tabular}\\
Remnant & $\rem$ & Subjects with covariate ($\bm{x}$) and outcome ($Y$) data available, but who were randomized between conditions in the RCT & n/a \\
\hline
\end{tabular}

\label{tab:sets}
\end{table}

Consider an A/B test in which subjects $i=1,\dots,n$ are randomized between two conditions, which we denote as $Z_i=0$ or $Z_i=1$, with the goal of estimating effects of $Z_i$ on an outcome $Y_i$.
Call the set of randomized subjects $i$ the ``RCT set,'' or $\rct$.
Typically, researchers running A/B tests are interested in the effect of $Z$ on a broader population than 
$\rct$, such as all users of the system, or all users of a particular type; denote this target population as $\pop$. 
For instance, students in a set of participating classrooms ($\rct$), working on a mastery-based homework assignment, may be randomized to receive tutoring in the form of either multi-step hints ($Z=1$) or complete explanations of problem solutions ($Z=0$), with the ultimate goal of estimating the effects of hints versus explanations on assignment completion ($Y$) for all users of the educational software ($\pop$). 
(We focus on binary treatments for the sake of simplicity, though the methods and concepts we discuss extend easily to experiments with more than two conditions.) 

Following \cite{neyman,rubin} let $y_i(z)$, $z=0,1$ represent the outcome that subject $i$ would experience if randomized to $z$---that is, if $Z_i=0$, the observed outcome $Y_i=y_i(0)$, and if $Z_i=1$ then $Y_i=y_i(1)$. 
Then, define the treatment effect for subject $i$ as $\tau_i=y_i(1)-y_i(0)$, the difference between the outcome $i$ would experience under condition 1 versus what they would experience under condition 0.

The challenge of causal inference is that for each $i$, only one of $y_i(0)$ or $y_i(1)$ is observed.
Hence, individual treatment effects $\tau_i$ cannot be estimated directly (at least, not precisely), but under some circumstances, average treatment effects can be estimated. 

\subsubsection{The Sample Average Treatment Effect}
First, consider the sample average treatment effect,
\begin{equation*}
    \sate=\sum_{i=1}^n \tau_i/n=\overline{\yt}-\overline{\yc},
\end{equation*}
where $\overline{\yt}$ is the sample average of $\yt$ over every subject in the RCT (whether $Z=1$ or $Z=0$), and $\overline{\yc}$ is the sample average of $\yc$. 
Hence, $\sate$ is never observed, but can often be estimated.
Claims about $\sate$ pertain only to the participants in $\rct$, not (necessarily) to treatment effects among other subjects. 

\subsubsection{The Population Average Treatment Effect}
When researchers' interest goes beyond the average effect in $\rct$, and actually pertains to the larger population $\pop$, then the estimand of interest is the population average effect, denoted $\pate$\footnote{We use the notation of expected value $\EE[\cdot]$ instead of the sample average $\bar{\cdot}$ since it will often be mathematically convenient to think of $\pop$ as an infinite ``super-population'' from which subjects are drawn randomly (see, e.g. \cite{ding2017bridging}).}.
If $\rct$ is a random sample of $\pop$, then there is little difference between estimating $\sate$ and estimating $\pate$. 
However, it is often the case that experimental participants are not representative of $\pop$.

\subsubsection{Subgroup Effects}
If prior to randomization the population is partitioned into $K$ subgroups---for instance, students with high or low prior academic performance, students in different school districts, or students in different demographic categories---then let $G_i\in 1,\dots, K$ denote subject $i$'s group membership, so that if $G_i=k$, then $i$ is in the $k^{th}$ subgroup. 
Then $\sate[G=k]$ and $\patesub{G=k}$ are average treatment effects for members of the subgroup $k$ in $\rct$ or the population $\pop$, respectively. 
In general, $\sate=\sum_{k=1}^K p_k \sate[G=k]$ and $\pate=\sum_{k=1}^K \pi_k \patesub{G=k}$, where $p_k$ and $\pi_k$ are the proportions of $\rct$ and $\pop$, respectively, that belonging to group $k$.

\subsection{Estimation and Types of Causal Bias}\label{sec:bias}
Bias in estimating average effects depends on the causal estimand of interest and can be due to bias in estimating $\sate$, which we will call ``internal" bias, bias in estimating $\pate$ due to differences between subjects in the experiment and the population, which we will call ``external bias,'' or a combination of the two. 
Our terminology mirrors the distinction between internal and external validity \cite[for eg.]{mcdermott2011internal}. 

\subsubsection{Aside: Why do We Care about Statistical Bias?}
While a good amount of early work in theoretical statistics focused on unbiased estimators, recent decades have seen increasing acknowledgment that unbiased estimators are often sub-optimal according to alternative estimation criteria and that a small amount of statistical bias may be a reasonable price to pay for improved statistical precision. 
That being the case, what accounts for our focus on unbiased estimation in this paper?

Although exact unbiasedness may not be an important goal for estimation in general, the concept of bias remains a useful formalization of some very important problems in estimation. 
For instance, the widely-known problems of estimating population quantities from unrepresentative or non-random samples or estimating causal effects from observational studies with unobserved confounding variables are both---in our opinion---most easily and clearly expressed in terms of bias. 
Extrapolation from unrepresentative samples and confounding can cause estimators to be inconsistent or inadmissible, and for confidence intervals and hypothesis tests to under-cover or over-reject, respectively.
Our focus is on bias since we take it to be the simplest and most straightforward way to formalize confounding and unrepresentative sampling.

\subsubsection{Estimating $\sate$ and Internal Bias}\label{sec:internal-bias}
In a completely randomized experiment, the set of subjects with $Z=1$ are a random sample of all the experimental participants, so $\overline{Y}_{Z=1}=(\sum_{i=1}^n Y_iZ_i)/(\sum_{i=1}^n Z_i)$, the average observed outcome for treated subjects, is an unbiased estimate of $\overline{\yt}$, and likewise $\overline{Y}_{Z=0}$ is an unbiased estimator of $\overline{\yc}$. 
(In general, let $\overline{X}_G$ be the sample mean of $X$ for subjects for whom $G$ is true $(\sum_{i=1}^n X_i\mathbf{1}\{G_i\})/(\sum_{i=1}^n \mathbf{1}\{G_i\})$, where $\mathbf{1}\{G_i\}=1$ if $G$ is true for $i$ and 0 otherwise.)
Then 
\begin{equation*}
    \tdm=\overline{Y}_{Z=1}-\overline{Y}_{Z=0},
\end{equation*}
the ``difference-in-means" or ``T-Test'' estimator, is (internally) unbiased for $\sate$.
The same reasoning extends to estimates of subgroup effects $\sate[G=k]$---the difference in mean outcomes for RCT subjects with $G=k$ between $Z=1$ and $Z=0$ is internally unbiased, i.e. unbiased for $\sate[G=k]$.

However, if treatment $Z$ is not randomized---or if randomization is ``broken'' due to attrition or some other irregularity---then $\tdm$ will be biased due to confounding. 
Similarly, if treatment was randomized, but with different probabilities of treatment assignment for different subjects, $\tdm$ may be a biased estimate of $\sate$. 

Even in a completely randomized experiment without other complications, some common effect estimators are biased for $\sate$.
For instance, say a vector of covariates $\bm{x}_i$ is observed for each subject.
The \textsc{ancova} estimator for $\sate$, the estimated coefficient on $Z$ from an ordinary least squares (\textsc{ols}) regression of $Y$ on $Z$ and $\bm{x}$, is biased for $\sate$ unless the linear model is correct.
In general, non-linear relationships between $\bm{x}$ and $y(0)$ or $y(1)$, or un-modeled interactions between the treatment indicator and $\bm{x}$, will lead to bias in the \textsc{ancova} estimator. 
That said, when $\bm{x}$ has low dimension relative to $n$, the bias of the \textsc{ancova} estimator is negligible (under suitable regularity conditions it decreases roughly with $1/n$; \citeN{freedman}).
However, if $\bm{x}$ has high dimension relative to $n$,
or if a prediction algorithm other than \textsc{ols} is used (improperly),  the bias might be substantial. 

\subsubsection{Estimating $\pate$ and External Bias}\label{sec:patebias}
An unbiased estimator of $\sate$ may still be biased for $\pate$, depending on the population of interest $\pop$.
For instance, consider a stylized example in which $G$ encoded income level: poor $G=1$ versus rich $G=2$, and that the effect of an intervention differs by income level---say $\patesub{G=1}<\patesub{G=2}$---and that sample proportions $p_1<p_2$ while population proportions $\pi_1>\pi_2$, so the experiment was conducted among subjects who were wealthier, on average, than the population of interest. 
Finally, say that within income groups $G$, the experimental subjects are representative of the corresponding subgroups in the population so that $\EE[\sate[G=k]]=\patesub{G=k}$.
Let $\hat{\tau}$ be an unbiased estimator of $\sate$.
As an estimate of the population average effect $\pate$, $\hat{\tau}$ will be biased:
\begin{equation}\label{eq:popbias1}
\begin{split}
&\EE[\hat{\tau}]-\pate=\EE[\sate]-\pate \\
=&p_1\EE\left[\overline{\tau}_{G=1}\right]+(1-p_1)\EE\left[\overline{\tau}_{G=2}\right]
-\pi_1\patesub{G=1}-(1-\pi_1)\patesub{G=2} \\
=&(p_1-\pi_1)\patesub{G=1}+(\pi_1-p_1)\patesub{G=2}\\
=&(p_1-\pi_1)(\patesub{G=1}-\patesub{G=2})>0
\end{split}
\end{equation}
since $p_2=1-p_1$ and $\pi_2=1-\pi_1$.
It is clear from \eqref{eq:popbias1} that if either $p_1=\pi_1$, so that the subjects in the experiment are representative of $\pop$, or if $\patesub{G=1}=\patesub{G=2}$, so that the average effect of the treatment doesn't vary with $G$, that $\hat{\tau}$ will be unbiased. 
In general, for an estimate to be externally biased, there must be at least one (observed or unobserved) characteristic in which the subjects in the experiment do not represent the population, \emph{and} which predicts variation in the treatment effect. 
If the ways in which the experimental sample is unrepresentative are unrelated to treatment effect variation, then there will be no external bias. 

Since, in the example above, $\hat{\tau}$ was unbiased for $\sate$, the bias of \eqref{eq:popbias1} is purely external bias. 
However, if internal bias is also present, then the two biases add, so that 
\begin{equation}
    \EE\left[\hat{\tau}\right]-\pate=\text{internal bias}+\text{external bias}.
\end{equation}
Note, however, that if internal and external bias have opposite signs, they may (partially) cancel each other out---that said, it is hard to know when this fortunate situation may or may not hold. 

Sometimes subgroup effect estimates can be combined to mitigate external bias from an unrepresentative RCT sample, via post-stratification \cite{miratrix2013adjusting}.
Say $\EE[\sate[G=k]]\approx\patesub{G=k}$ as in the example above, and that population proportions $\pi_k$ are known.
Let $\hat{\tau}_k$ be unbiased estimates of $\sate[G=k]$; then,
\begin{equation}\label{eq:post-stratification}
\EE\left[\sum_k \pi_k \hat{\tau}_k\right]=\sum_k \pi_k \EE[\hat{\tau}_k]\approx \sum_k. \pi_k \patesub{G=k}=\pate
\end{equation}
That is, if estimated subgroup effects $\hat{\tau}_k$ are unbiased, then the post-stratification estimator $\sum_i \pi_k \hat{\tau}_k$ will also be externally unbiased.
Hence, accurate estimation of subgroup effects can reduce external bias of overall population effects.

\subsection{Internally Unbiased Estimators using Auxiliary Data}

\subsubsection{The Remnant}
While the difference-in-means estimator $\tdm$ is unbiased for $\sate$ in a completely randomized experiment, it may be imprecise, especially when the sample size is small.
This problem may be exacerbated if a researcher is interested in estimating subgroup effects, either because of scientific interest in subgroups or for the sake of post-stratification.
The reason is that $\sate$ depends on unobserved counterfactual potential outcomes, $y_i(0)$ if $Z_i=1$ and $y_i(1)$ if $Z_i=0$, which must be imputed.
$\tdm$ relies on very rudimentary imputation strategy: the imputed $\hat{y}_i(0)=\overline{Y}_{Z=0}$ for all $i$ such that $Z_i=1$, and $\hat{y}_i(1)=\overline{Y}_{Z=1}$ for all $i$ such that $Z_i=0$. 
This strategy ignores all observed differences between subjects in the experiment, instead imputing one of the same two values for every subject. 

In many cases, covariate and outcome data from an experiment are drawn from a larger database.
For instance, educational field trials may use state longitudinal data systems to collect covariate data on student demographics and prior achievement as well as on outcomes of interest such as standardized test scores, and medical trials may gather baseline and outcome data from databases of medical records.
Most relevant for our purposes, analysis of A/B tests within online applications can access rich baseline data from users' logs prior to the onset of the experiment and often draw outcome data from that same source. 
In these cases, researchers have the option of gathering additional auxiliary data---covariate and outcome data from users who were not part of the experiment.
This includes historical data from before the onset of the experiment, as well as data from concurrent users who were not part of the experiment for some other reason. 
We refer to this set of users as the ``remnant'' from the experiment \cite{rebarEDM} (rounding out the list of sets described in Table \ref{tab:sets} and Figure \ref{fig:sets}). 

\subsubsection{A Naive Estimator using the Remnant}
Say, for the sake of argument, that every subject in the remnant was in the $Z=0$ condition; this will be the case if, for instance, $Z=0$ represents a ``business as usual'' condition.
Then, say researchers used the remnant to train an algorithm $\predr{\bm{x};\bm{\beta}}$ predicting outcomes from covariates $\bm{x}$, with parameters $\bm{\beta}$, estimated with remnant data as $\hat{\bm{\beta}}$.
Define this algorithm's prediction for each experimental subject $i$ as $\predri\equiv \predr{\bm{x}_i;\hat{\bm{\beta}}}$ (where ``$\equiv$'' denotes definition). 
Researchers could use these to impute control potential outcomes $y(0)$ for participants in the experiment as $\hat{y}_i(0)=\predri$. 
That is, for each experimental participant with $Z_i=1$, estimate an individual treatment effect of $\hat{\tau}_i=Y_i-\predri$ and estimate $\bar{\tau}$ or $\EE_{\pop}[\tau]$ as $\hat{\tau}_{naive}=\overline{\hat{\tau}}_{Z=1}$.

The estimator $\hat{\tau}_{naive}$ has the potential to be much more precise than $\tdm$ since it can account for observed baseline differences between experimental subjects, and use those differences to tailor its imputations to each individual subject. 
On the other hand, it has two serious disadvantages.
First, the participants in the experiment are not necessarily drawn from the same population as the remnant, so there is no guarantee that the conditional distribution of $\yc$ given $\bm{x}$ is the same in both groups. 
If the remnant is not representative of the experiment so that $p(\yc|\bm{x})$ differs between the two sets, $\hat{\tau}_{naive}$ may be biased for both $\bar{\tau}$ and $\pate$.
Second, even if the remnant is representative of the sample, there is typically no guarantee that the predictions $\predr{\bm{x};\bm{\beta}}$ are unbiased---in this case, the often erratic behavior of supervised learning algorithms in finite samples can also lead to bias. 

\subsubsection{Better Estimation using the Remnant}
Both of these disadvantages can be corrected by relying on \emph{both} randomization and supervised learning from the remnant. 
Specifically, the problems that cause internal bias in $\hat{\tau}_{naive}$ will also be present when comparing $Y_i$ to $\predri$ for subjects in the control group, leading to the ``remnant-based residualization'' or ``rebar'' estimator \cite{rebarEDM}, 
\begin{equation}\label{eq:rebar}
\rebar\equiv\hat{\tau}_{naive}-\overline{Y-\bm{\predri[]}}_{Z=0}=\overline{Y-\bm{\predri[]}}_{Z=1}-\overline{Y-\bm{\predri[]}}_{Z=0}=\tdm-\overline{\bm{\predri[]}}_{Z=1}-\overline{\bm{\predri[]}}_{Z=0},
\end{equation}
where $\bm{\predri[]}$ is the vector of imputations $\{\predri\}_{i=1}^n$. 
As \eqref{eq:rebar} suggests, there are (at least) two ways to conceptualize the rebar estimator: first, it corrects the bias of $\hat{\tau}_{naive}$ by subtracting the analogous contrast in the $Z=0$ group, $\overline{Y-\bm{\predri[]}}_{Z=0}$, and second, it corrects for imprecision in $\tdm$ by subtracting the finite-sample difference in $\bm{\predri[]}$ between students in the two treatment conditions. $\rebar$ is precise if $\bm{\predri[]}$ is close to $\yc$, on average, and is always unbiased for $\bar{\tau}$, due to the randomization of treatment assignment.
Importantly, because the parameters $\beta$ from the algorithm $\predr{\bm{x};\bm{\beta}}$ are estimated using a separate sample, and $\bm{x}$ is fixed at baseline, $\rebar$ will be unbiased for $\bar{\tau}$ regardless of whether imputations $\bm{\predri[]}$ are themselves accurate or biased. 
This property is guaranteed by the randomization of treatment assignment. 
In fact, it applies regardless of whether subjects in the remnant were in the $Z=0$ or $Z=1$ condition, or some other condition altogether. 

The problem with $\rebar$ is that if the algorithm $\predr{\bm{x};\bm{\beta}}$ performs poorly for subjects in $\rct$, then $\rebar$ will have high variance---sometimes even higher than $\tdm$.
A better solution is based on the fact that, in essence, $\predri$ is itself a covariate, since it is a function of covariates $\bm{x}_i$ and parameters $\bm{\beta}$ estimated using a separate sample. 
That being the case, it can be used as a covariate, perhaps along with others, in an existing covariate-adjusted estimator of $\sate$.

For instance, consider the \textsc{ancova} estimator based on the following \textsc{ols} model:
\begin{equation}\label{eq:ols}
Y_i = \alpha_0+\alpha_1 \predri[]+\tau_{OLS} Z_i+\epsilon_i,
\end{equation}
where $\epsilon_i$ is a mean-0 error term. 
The estimated coefficient on $Z$ from this model, $\hat{\tau}_{OLS}$, triangulates between $\rebar$ and $\tdm$, essentially picking whichever estimator is better.
If $\predri[]$ is highly correlated with $Y$, then we might expect its estimated coefficient in \eqref{eq:ols} $\hat{\alpha}_1\approx 1$, in which case $\hat{\tau}_{OLS}\approx \rebar$.
If, on the other hand, $\predri[]$ is a poor prediction of $Y$, then $\hat{\alpha}_1 \approx 0$ and $\hat{\tau}_{OLS}\approx \tdm$. 
However, as discussed earlier, while $\hat{\tau}_{OLS}$ is a consistent estimator of $\sate$, it is slightly biased, and its associated standard error estimates require either large samples or additional modeling assumptions. 
Researchers may consider including additional covariates as predictors alongside $\predri[]$ in an \textsc{ancova} model like \eqref{eq:ols}; however, as the number of covariates, interactions and/or non-linear terms increases, so may bias or other inferential issues. 

Alternatively, \citeN{gagnon2021precise} suggests incorporating $\bm{\predri[]}$, perhaps alongside other covariates, into a flexible, internally-unbiased effect estimator that adjusts for baseline covariates \cite[for eg.]{wager,aronowMiddleton}.
We will focus here on the ``\textsc{loop}'' estimator \cite{loop}.
As above, for each subject $i$, let $\bm{x}_i$ be a vector of covariates.
In general, \textsc{loop} is an alternative to \textsc{ancova} for A/B tests with Bernoulli randomization, in which each subject is independently randomized with $Pr(Z_i=1)=p$ for all $i$.
Like \textsc{ancova}, \textsc{loop} estimates $\sate$ after adjusting for baseline differences in $\bm{x}$ between subjects assigned to $Z=0$ and $Z=1$.
Unlike \textsc{ancova}, \textsc{loop} estimates are exactly unbiased for $\sate$ and are not limited linear models of the covariates---in principle, they can accommodate any model relating outcomes to covariates, including models that can incorporate high-dimensional covariate matrices. 
Write the standard \textsc{loop} estimator, adjusting for covariates $\bm{x}$, as $\myloop$---this estimator adjusts for $\bm{x}$ but does not use the model trained in the remnant. 
We recommend two alternatives: $\reloopV$, which adjusts the estimate for $\bm{\predri[]}$ instead of $\bm{x}$---this estimator is quite similar to $\hat{\tau}_{OLS}$ in \eqref{eq:ols} but unbiased---and $\reloop$, which adjusts for \emph{both} $\predr{\bm{x}}$ and $\bm{x}$, incorporating all of the best properties of $\tdm$, $\myloop$, and $\rebar$.
These estimators are design-based, exactly unbiased for the $\sate$, give conservative standard error estimates, and make no modeling assumptions, beyond the design of the experiment itself.

The following sub-section gives a more technical description of the \textsc{loop} estimator and $\reloop$ for interested readers.

\subsubsection{The \textsc{loop} Estimator with Remnant-Based Predictions}

In a Bernoulli-randomized A/B test, specify an algorithm $\pred{z}{\bm{x},\predri[] ;\bm{\alpha}}$ to impute potential outcomes $\yc$ and $\yt$ from remnant-based imputations $\bm{\predri[]}$, and (optionally) covariates $\bm{x}$, with parameters $\bm{\alpha}$. 
(Note that there are two separate algorithms predicting $Y$ from $\bm{x}$: $\predr{\bm{x};\bm{\beta}}$ is fit using data from the remnant and produces imputations $\predri$, while $\pred{z}{\bm{x},\predri[] ;\bm{\alpha}}$ is fit using RCT data.)
For instance, \cite{gagnon2021precise} considers models 
\begin{equation}\label{eq:olsAdj}
\pred{z}{\predri[] ;\bm{\alpha}}_{OLS}=\alpha_0^z+\alpha_1^z\predri[],
\end{equation}
where $\bm{\alpha}=[\alpha_0,\alpha_1]$, an \textsc{ols} intercept and slope estimated separately in each treatment arm, as well as a random forest (RF) predictor, $\pred{z}{\bm{x},\predri[] ;\bm{\alpha}}_{RF}$ incorporating covariates $\bm{x}$ alongside $\bm{\predri[]}$ as predictors, but ultimately recommends an ensemble of the two. 

To estimate $\sate$ without bias, it is essential that the predictions from $\pred{z}{\bm{x},\predri[];\bm{\alpha}}$ be statistically independent from the treatment assignment $Z$.
The recommended estimators in \cite{gagnon2021precise} ensure that this is the case by using leave-one-out sample-splitting.
For each subject in the experiment $i=1,\dots,n$, estimate $\bm{\alpha}$ using data from the other $n-1$ subjects. 
Denote this estimate of $\bm{\alpha}$, using data from all $\rct$ subjects except $i$, as $\hat{\bm{\alpha}}_{(i)}$. 
Impute missing potential outcomes using predictions $\predic(\predri[],\bm{x})=\pred{0}{\predri,\bm{x}_i;\hat{\bm{\alpha}}_{(i)}}$ and $\predit(\predri[],\bm{x})=\pred{1}{\predri,\bm{x}_i;\hat{\bm{\alpha}}_{(i)}}$.

Finally, estimate $\sate$: first, let $\hat{m}_i(\predri[],\bm{x})=p\predic(\predri[],\bm{x} )+(1-p)\predit(\predri[],\bm{x})$, an imputation of $i$'s expected counterfactual potential outcome.
Then estimate $\overline{\tau}$ as:
\begin{equation}\label{eq:loop}
    \reloop=\sum_{i:Z_i=1}\frac{Y_i-\hat{m}_i(\predri[],\bm{x})}{np}- \sum_{i:Z_i=0}\frac{Y_i-\hat{m}_i(\predri[],\bm{x})}{n(1-p)},
\end{equation}
where $p$, as above, is the probability of an individual participant being assigned to the $Z=1$ condition. 
The $(\bm{\predri[]},\bm{x})$ in the notation $\reloop$ refer to the data included in the imputation algorithms that give rise to $\hat{m}$. In the following section, we contrast $\reloop$ with variants $\reloopV$, in which $\hat{m}_i=\hat{m}(\predri)$ is a function of $\predri$ only, and $\myloop$, in which $\hat{m}_i=\hat{m}(\bm{x}_i)$ is a function of $\bm{x}_i$ but not $\predri$. 

$\reloop$ and its variants are inverse-probability-weighted estimates (also called Horvitz Thompson)---they are similar in form to $\tdm$, except with the treatment and control sample sizes replaced with their expected values, $np$ and $n(1-p)$.
Aside from that difference, $\reloop$ with $\hat{m}_i=0$ would correspond to $\tdm$, and $\reloopV$ with $\hat{m}_i=\predri$ would be equivalent to $\rebar$.
In general, $\reloop$ is much more flexible than either $\tdm$ or $\rebar$, since it allows $\predri[]$'s role to vary depending on its prognostic value, and because it allows flexible incorporation of other baseline covariates $\bm{x}$. 

Because parameters $\bm{\alpha}$ are estimated independently of $i$'s outcome data, and $\bm{x}_i$ is fixed prior to treatment assignment, the sample splitting estimator is unbiased for the sample average treatment effect $\overline{\tau}$. 

In \citeN{gagnon2021precise}, incorporating $\predri$ into the \textsc{loop} estimator led, in many cases, to substantial gains in precision compared to either $\tdm$ or to the \textsc{loop} estimator with other covariates but not $\predri$.

None of the methods considered here assumes that either imputation model, $\alg$ or $\pred{z}{\bm{x},\predri[] ;\bm{\alpha}}$ is correct, unbiased, or consistent in any sense. 
Regardless of the quality of the imputation methods, randomization of treatment assignment ensures that effect estimates are unbiased.

\subsubsection{Specific Estimators and Associated Terminology}\label{sec:estimators}

Our two recommended estimators, which we term \textsc{r}e\textsc{loop} and \textsc{r}e\textsc{loop}+, combine ideas from $\rebar$ \eqref{eq:rebar} and the leave-one-out covariate adjustment strategy \textsc{loop} \cite{loop}---hence the name ``\textsc{r}e\textsc{loop}.'' We will compare \textsc{r}e\textsc{loop} and \textsc{r}e\textsc{loop}+ to the T-Test estimator $\tdm$, and a \textsc{loop} estimator that does not use remnant data. All told, we consider four different estimators:
\begin{itemize}
\item ``T-Test": the difference-in-means estimator $\tdm$, with no covariate adjustment
\item ``\textsc{loop}'': $\myloop$ adjusts for covariates using a random forest imputation model fit to $\rct$ data. It does not use any remnant data. 
\item ``\textsc{r}e\textsc{loop}'': $\reloopV$ adjusts only for $\predri$, imputations from the model trained in the remnant, using \textsc{loop} with the \textsc{ols} $\rct$ imputation model \eqref{eq:olsAdj}. It adjusts for no other covariates.
\item ``\textsc{r}e\textsc{loop}+'': $\reloop$ uses an ensemble of \textsc{ols} and random forests trained in $\rct$ to adjust for both $\predri$ and other covariates. 
\end{itemize}

When an imputation model $\pred{z}{\bm{x},\predri[] ;\bm{\alpha}}$ is trained using RCT data, we refer to the associated covariate adjustment as ``within-sample'' or ``within-RCT'' adjustment.
When an imputation model is trained in the remnant (i.e. $\predr{\bm{x};\bm{\beta}}$), we refer to the associated covariate adjustment as ``remnant-based.''
Comparing the two types of adjustment, within-sample adjustment has the advantage of hewing more closely to the actual $\rct$ data on which it's trained, while remnant-based adjustment can rely on models fit using the remnant, which may boast a much larger sample size than the $\rct$. 
\textsc{r}e\textsc{loop} and \textsc{r}e\textsc{loop}+ make use of both types of adjustment. 

\subsubsection{Estimating Sampling Variance, p-values, and Confidence Intervals}

The true sampling variances of $\tdm$,  $\rebar$, and 
$\reloop$, as estimates of $\sate$, depend on the correlation between $\yc$ and $\yt$, which is not identified without making further assumptions, since $\yc$ and $\yt$ are never observed simultaneously. 
However, it is possible to \emph{conservatively} estimate the sampling variances of all three estimators.
Specifically, for $z=0,1$, let 
$$\hat{E}^2_z=\frac{1}{n_z}\sum_{i:Z_i=z}\left[\pred{z}{\predri,\bm{x}_i;\hat{\bm{\alpha}^z}_{(i)}}-Y\right]^2.$$
Then estimate the sampling variance of $\reloop$ as:
$$\widehat{\VV}\left(\reloop\right)\frac{1}{n}\left[\frac{p}{1-p}\hat{E}^2_0+\frac{1-p}{p}\hat{E}^2_1+2\hat{E}_0\hat{E}_1\right].$$
As \citeN{loop} show, $\EE\left[\widehat{\VV}\left(\reloop\right)\right]\ge \VV\left(\reloop\right)$---that is, $\reloop$'s estimated sampling variance is conservative in expectation.

Let the estimated standard error of $\reloop$ $\widehat{SE}=\widehat{\VV}\left(\reloop\right)^{1/2}$. 
The usual $1-\alpha$ confidence interval has asymptotic coverage of at least $1-\alpha$---i.e.
$$Pr(\bar{\tau}\in \reloop\pm \mathfrak{z}_{1-\alpha/2}SE)\rightarrow 1-\tilde{\alpha} \ge 1-\alpha$$ 
as $n\rightarrow \infty$, where $\mathfrak{z}_{1-\alpha/2}$ is the $1-\alpha/2$ quantile of the standard normal distribution.
Similarly, a hypothesis test that rejects the null hypothesis of $\bar{\tau}=0$ when $|\reloop/SE|\ge \mathfrak{z}_{1-\alpha/2}$ will have a type-I error rate of at most $\alpha$ in large samples. 

The possible upward bias in these variance estimates will, if anything, cause confidence intervals to include the true parameter too often, or cause type-I error rates to be too low. 
While an unbiased sampling variance estimator would be preferable, conservative estimators are (arguably) the next best thing.

\section{Data from 68 Educational A/B Tests}\label{sec:data}
The remainder of the paper will discuss a set of illustrations and case-studies in using the \textsc{r}e\textsc{loop} and \textsc{r}e\textsc{loop}+ to estimate causal treatment effects from A/B tests run on an educational technology platform. 
This section describes the dataset---first the A/B tests themselves, and then the remnant---and the following section describes $\predr{\bm{x};\bm{\beta}}$, the deep-learning model trained using remnant data. 
Subsequent sections will use data from the A/B tests and imputations from the model trained in the remnant to answer our research questions.

 E-Trials is a platform that allows researchers to design educational experiments that will then be run within the ASSISTments online tutor. Education researchers can specify experimental conditions, including  variation on how subject matter is portrayed, available hints, and feedback to students. Researchers also choose learning modules on which their experiments run.
 When teachers subsequently assign these modules to their students, the students are randomized between the conditions.
After the period of the experiment has ended, the researcher is provided with a dataset, including classroom and student identifiers, log data from during the experiment, and outcome data such as which students completed the assignment and how many problems they worked.
 Students are randomized between conditions independently, one at a time; when there are only two conditions, this is Bernoulli randomization.

We gathered data from a set of 84 A/B tests run on E-Trials. Since our interest here is primarily methodological, with the goal of reducing standard errors, we focus on estimated standard errors as opposed to treatment effects. Our analyses focus on assignment completion as a binary outcome.

We also gathered a set of nine student-level aggregated predictors, to be used for within-RCT covariate adjustment. These were the numbers of skill builders (mastery-based modules in ASSISTments) and problem sets each student began and completed, as well as each student's prior median first response time when working ASSISTments problems, median time on task, overall correctness, completed problem count, and average attempt count.   

Several experiments included multiple conditions, rather than only treatment and control. We assume that the primary interest in these experiments lies in head-to-head comparisons between conditions, and, as such, we analyze all unique pairs of conditions within randomized experiments separately.
All in all, this included 383 pairs.
However, not every pair was amenable to analysis. Six pairwise contrasts were dropped because the outcome variance in one or both of the conditions was zero.
Further exclusions were motivated by two factors: first, the \textsc{loop} estimator (which also underlies the \textsc{r}e\textsc{loop} and \textsc{r}e\textsc{loop}+ estimators) presumes that $p_i=Pr(Z_i=1)$ is known. When the experiments were run, the E-Trials platform was only equipped to run Bernoulli-randomized experiments in which students were independently assigned to available conditions with equal probability. Hence, in theory, $p=1/2$ should hold in all pairwise comparisons. However, there were strong indications that a handful of experiments used a different randomization scheme---we suspect that in some cases two conditions were combined, leading to $p=2/3$ or 1/3.
To exclude cases in which $p\ne 1/2$, we estimated p-values testing the null hypothesis that $p=1/2$ for each comparison we considered; we dropped contrasts in which the p-value testing $p=1/2$ was $<0.1$.
Secondly, there were some contrasts that included extremely small samples, with the smallest being $n=16$.
The \textsc{loop} estimators rely on \textsc{ols} regression or more complex models, and cannot be expected to perform well when sample sizes are so small.
In the main analyses, we dropped experiments in which the sample size in either condition was less than $5(k+2)+1$, where $k=9$ is the number of predictors, which would allow for at least 5 observations per predictor in any model.
In the subgroup analyses of Section \ref{sec:subgroup}, we analyzed subgroups with smaller sample sizes.

These exclusions left a total of 227 randomized contrasts---pairs of treatment conditions between which students were randomly assigned---drawn from 68 separate A/B tests. 

\subsection{Data Collection}

%

The data was collected from ASSISTments in two sets: remnant data and experiment data. Remnant data was used to train the imputation models, and experiment data was used to impute the outcomes in each experiment using the imputation models. The skill builders started by the students in the remnant data were not the same skill builders as the experimental skill builders in the experiment data, nor is there any overlap in students between the two datasets. \textbf{No information from the students or skill builders in the experiment data was in the remnant data used to train the imputation models}.

For both the remnant and experiment data, the same information was collected. For each instance of a student starting a skill builder for the first time, we collected data on whether they completed the skill builder, and if so, how many problems they had to complete before mastering the material. The imputation models, discussed more in section \ref{sec:imputation}, were trained to predict these two dependent measures. The data used to predict these dependent measures was aggregated from all of the previous work done by the student. Three different sets of data were collected for each sample in the datasets: prior student statistics, prior assignment statistics, and prior daily actions. Prior student statistics included the past performance of each student, for example, their prior percent correct, prior time on task, and prior assignment completion percentage. Prior assignment statistics were aggregated for each assignment the student started prior to the skill builder. Prior assignment statistics included things like the skill builders' unique identifiers (or in the remnant data, the ID of the experimental version of a skill builder, if it existed), how many problems had to be completed in the assignment, students' percent correct on the assignment, and how many separate sessions students used to complete the assignment. Prior daily actions contained the total number of times students performed each possible action in the ASSISTments tutor for each day prior to the day they started the skill builder. The possible actions included things like starting a problem, completing an assignment, answering a problem, and requesting support. Complete lists of features included in prior student, assignment, and daily action datasets are included in Tables \ref{tab:pssf}, \ref{tab:pasf}, and \ref{tab:pdaf} in the appendix. 193,218 sets of prior statistics on students, 837,409 sets of statistics on prior assignments, and 695,869 days of students' actions were aggregated for the remnant data, and 113,963 sets of prior statistics on students, 2,663,421 sets of statistics on prior assignments, and 926,486 days of students' actions were aggregated for the experiment data.

\section{Remnant-Trained Imputation Models}\label{sec:imputation}

\subsection{Model Design}

\begin{figure}
\includegraphics[width=\textwidth]{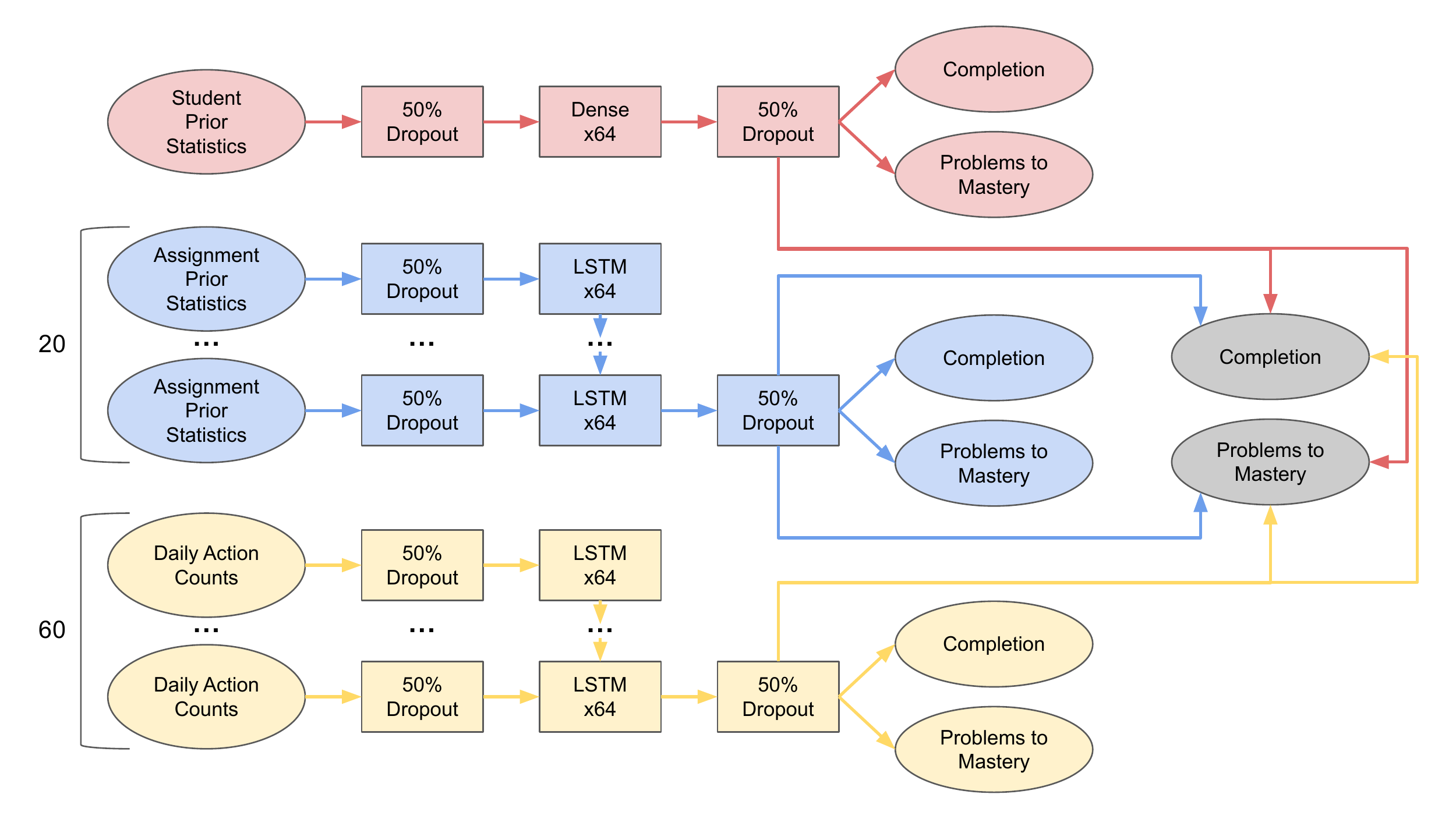}
\caption{All four of the imputation models in one. The red model predicts performance using only prior statistics of the student, the blue model uses statistics on the last 20 assignments completed by the student to predict performance, and the yellow model uses the last 60 days of actions the student took in the tutor. The combined model, shown in grey, uses all three models to predict performance.}
\label{fig:models}
\end{figure}

Each of the three types of data in the remnant dataset was used to predict both skill builder completion and the number of problems completed prior to demonstrating mastery. For each type of data: prior student statistics, prior assignment statistics, and prior daily actions, a separate neural network was trained. Additionally, a fourth neural network was trained using a combination of the previous three models. The prior student statistics model, shown in Figure \ref{fig:models} in red was a simple feed-forward network with a single hidden layer of nodes using sigmoid activation and dropout. Both the prior assignment statistics model and the prior daily actions model, shown in Figure \ref{fig:models} in blue and yellow respectively, were recurrent neural networks with a single hidden layer of LSTM nodes \cite{gers2000learning} with both layer-to-layer and recurrent dropout. The prior assignment statistics model used the last 20 started assignments as input, and the prior daily actions model used the last 60 days of actions as input. The last 20 assignments were chosen because of success in prior work with similar numbers of prior assignments \cite{rebarEDM}, and the last 60 days of actions were chosen based on usage data that indicated that after two months, students are unlikely to be working on content that is relevant for predicting their assignment completion. The combined model in Figure \ref{fig:models} takes the three models above and couples their predictions such that the prediction is a function of all three models' weights and the loss is backpropagated through each model during training. The hyperparameters of the model, including the dropout frequency, layer depth, and the number of nodes in each layer, were determined via grid search prior to using the model in the \textsc{r}e\textsc{loop} process. This model was chosen because this set of hyperparameters led to the lowest loss when predicting a held-out subset of the data.

Dropout was used to regularize model training but was not used in model validation, testing, or prediction. 

\subsection{Model Training}

To select the best model hyperparameters and to measure the quality of each imputation model, 5-fold cross-validation was used to train and calculate various metrics for each model. For all training, the ADAM method \cite{kingma2014adam} was used during backpropagation, binary cross-entropy loss was used for predicting completion, and mean squared error loss was used for problems to mastery. The total loss for each model was the sum of the two individual losses. Because mean squared error and binary cross-entropy have different scales, a gain of 16 was applied to the binary cross-entropy loss, which brought the loss into the same range as the mean squared error loss for this particular dataset. The gain of 16 was determined via grid search based on which gain led to the most accurate completion predictions during cross-validation because assignment completion was the outcome of interest for experiment analysis. Table \ref{tab:training} shows various metrics of the models' quality. Interestingly, even though all the models are bad at predicting problems to mastery, removing problems to mastery from the loss function reduced the models' ability to predict completion.

\begin{table}
\caption{Metrics Calculated from 5-Fold Cross Validation for each Model.}
\begin{tabular}{|r|c|c|c|c|}
\hline
       & Prior Student    & Prior Assignment & Prior Daily   & \\
Metric & Statistics       & Statistics       & Action Counts & Combined \\
\hline
Completion AUC       & 0.743 & 0.755 & 0.658 & \textbf{0.770} \\
Completion Accuracy  & 0.761 & 0.767 & 0.743 & \textbf{0.774} \\
Completion $R^2$     & 0.143 & 0.161 & 0.045 & \textbf{0.184} \\
\# of Problems MSE   & 8.489 & 8.505 & 8.719 & \textbf{8.363} \\
\# of Problems $R^2$ & 0.033 & 0.032 & 0.007 & \textbf{0.048} \\
\hline
\end{tabular}
\label{tab:training}
\end{table}

Based on Table \ref{tab:training}, statistics on prior assignments were the most predictive of students' assignment performance, followed by the students' overall prior performance statistics, and then their daily action history, which was the least predictive of their performance on their next assignment. Combining these datasets together led to predictions of a higher quality than any individual dataset could achieve. 

The effort we put into optimizing the model likely contributed to our methods' successes. 
However, our methods do not assume that the imputation model is optimal, accurate, or correct in any sense. 
A well-fitting model will lead to precise effect estimates, but estimates using a poorly-fitting model will still be unbiased, and their associated statistical inference will still be valid.

\section{Research Question 1: Can Imputations from Remnant-Trained Models Improve Standard Errors for Average Effects?}\label{sec:rq1}

To gauge the potential of remnant-based imputations to improve the precision of impact estimates, we compared estimated sampling variances from the four different treatment effect estimators listed in Section \ref{sec:estimators}: T-Test ($\tdm$), which includes no covariate adjustment; \textsc{loop}, which uses random forests for within-sample covariate adjustment using only the nine student-aggregated covariates in Section \ref{sec:data} but not the remnant; \textsc{r}e\textsc{loop}, which uses remnant-based imputations $\predri$ in a within-sample \textsc{ols} adjustment model; and \textsc{r}e\textsc{loop}+, which uses an ensemble algorithm to adjust for both $\predri$ and the nine student-aggregate covariates in \textsc{loop}. 
In this analysis, we used the ``combined" model, including all available remnant data, to generate remnant-based imputations $\predri$. 
We used these four estimators to estimate $\sate$ in each of the 227 randomized contrasts described above.

\begin{figure} 
\includegraphics[width=0.9\textwidth]{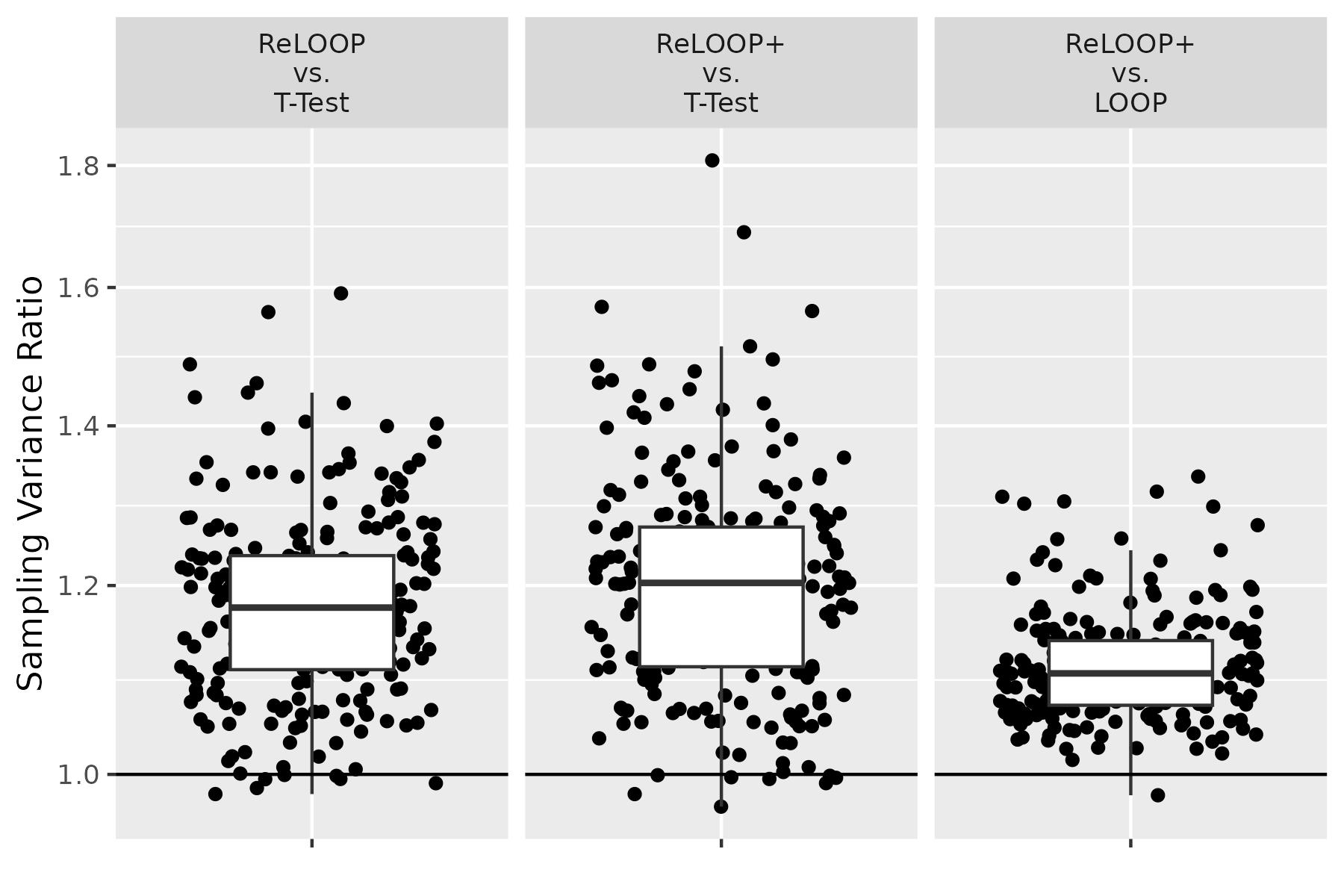}
\caption{Boxplots and jittered scatter plots of the ratios of estimated sampling variances of $\tdm$ (i.e. ``T-Test," which includes no covariate adjustment), $\myloop$ (``\textsc{loop}'', which adjusts for covariates within sample, but does not use the remnant), $\reloopV$ (``\textsc{r}e\textsc{loop},'' which adjusts for remnant-based imputations but not within-sample covariates), and $\reloop$ (``\textsc{r}e\textsc{loop}+," which adjusts for both within-sample covariates and remnant-based imputations) in 227 randomized contrasts. The Y-axis is on a logarithmic scale, so that, say, doubling the sample size appears as the same magnitude of an effect as halving the sample size.}
\label{fig:main}
\end{figure}

Figure \ref{fig:main} shows the ratios of estimated sampling variances from the four estimators. 
Since sampling variance scales as $1/n$, ratios of sampling variances can be thought of as ``sample size multipliers''---that is, decreasing the variance by a factor of $q$ is analogous to increasing the sample size by the same factor. 
The results in Figure \ref{fig:main} were previously reported in a conference poster \cite{sales2022more}. 

The panel on the left of \ref{fig:main} compares $\reloopV$ to $\tdm$, the T-Test estimator. In nearly every case the estimator using remnant data substantially outperformed the t-test estimator. In the majority of cases, including remnant-based predictions was roughly equivalent to increasing the sample by between 15 and 60\%.
The middle panel of Figure \ref{fig:main} compares $\tdm$ to $\reloop$. Here the results are slightly more impressive than those of the left panel---the median improvement is equivalent to increasing the sample size by about 20\%, and in the best case the improvement is equivalent to an 80\% increase in sample size. 

The rightmost panel of Figure \ref{fig:main} compares $\myloop$, which uses leave-one-out sample splitting and a random forest to adjust for covariates---but does not use the remnant---to $\reloop$ which does. In this case, we see more modest relative gains, which is to be expected since $\myloop$ can accomplish a good deal of covariate adjustment using only experimental data. 
Nevertheless, the contribution of the remnant is still significant---in roughly half of cases, including data from the remnant was equivalent to increasing the sample size by about 10--20\%, and in a handful of cases the improvement was closer to 30\%.

In summary, covariate adjustment can lead to substantial gains in precision, with the greatest improvement resulting from adjustment using both within-sample aggregated covariates and remnant-based imputations. In particular, estimators including remnant-based imputations consistently outperformed those that used only within-sample covariate adjustment. 

\subsection{Did the Remnant Help Us Discover any Effects?}\label{sec:full-p-values}
Researchers may naturally want to know if our claim to increase the power of A/B tests to detect effects actually lead, in practice, to more effects detected. 
In other words, did covariate adjustment lead to any p-values dipping below the $\alpha=0.05$ threshold?
Counting significant p-values is a problematic approach to gauging the success of our method since it depends on the size of the true effects. 
In particular, if the true $\sate$ is equal to 0, then a p-value less than 0.05 would be a type-I error, but if the $\sate$ is not equal to 0, a p-value less than 0.05 would be a true discovery. 
Since the ground truth is unknown, we cannot know which one is the case.

\begin{table}[ht]
\centering
\caption{The number of p-values less than $\alpha=0.05$ using each of the four estimators. The table counts significant p-values unadjusted for multiple comparisons, and adjusted with the Benjamini-Hochberg and Benjamini-Yekutieli procedures.}\vspace*{1ex}
\input{fullSignificance}

\label{tab:significanceFull}
\end{table}

Nevertheless, we will press on.
Table \ref{tab:significanceFull} gives the count of significant p-values using each of the four estimates. The first row gives a count of unadjusted p-values; if each pairwise comparison were considered in isolation, these would be the relevant counts. A researcher using T-Tests would report discoveries in 38 cases, researchers using  \textsc{loop} would report one additional discovery, and those using \textsc{r}e\textsc{loop} or \textsc{r}e\textsc{loop}+ would report an additional 3 discoveries.
However, since there were 227 total hypothesis tests, even if the null hypothesis were true in every case we would expect around 11 significant p-values; in other words, since we are considering the p-values as a group a multiplicity adjustment is in order. 
We considered two adjustment methods, both designed to limit the ``false discovery rate''---the proportion of the discoveries that are, in fact, type-I errors---to 5\%. 
The second row of Table \ref{tab:significanceFull} counts p-values adjusted with the Benjamini-Hochberg procedure \cite{benjamini1995controlling}. This procedure is guaranteed to control the false discovery rate only if the tests are independent\footnote{There are some types of dependence which are OK, too, but they are difficult to describe, much less to verify.}. The pairwise comparisons we consider are not independent, since each A/B test may have contributed several pairwise comparisons, which share data. 
After Benjamini-Hochberg adjustment, a researcher using T-Tests would only discover 3 effects, while researchers using \textsc{loop} would discover 11, those using \textsc{r}e\textsc{loop} would discover 8, and those combining within-sample and remnant-based adjustments with \textsc{r}e\textsc{loop}+ would lead to two additional discoveries or 10 total. 

It may be surprising that although \textsc{r}e\textsc{loop}+ standard errors tend to be smaller than those from \textsc{loop}, \textsc{loop} leads to one more discovery than \textsc{r}e\textsc{loop}+. In fact, there were two cases in which \textsc{loop} p-values were significant after Benjamini-Hochberg adjustment, but \textsc{r}e\textsc{loop}+ p-values were not. In both cases, \textsc{r}e\textsc{loop}+ standard errors were lower, but \textsc{r}e\textsc{loop}+ estimates were closer to zero as well. There was one case in which \textsc{r}e\textsc{loop}+ led to a significant p-value but \textsc{loop} did not; in this case, \textsc{r}e\textsc{loop}+ returned a smaller standard error and an effect estimate larger magnitude than \textsc{loop}.

The third row of the table counts significant p-values adjusted by the more conservative Benjamini-Yekutieli procedure \cite{benjamini2001control}, which controls the false discovery rate even under arbitrary dependence of tests. 
Researchers using any of the four estimators we've considered and adjusting with the Benjamini-Yekutieli procedure would all reject 2 null hypotheses among the 227 possibilities.

\subsection{Which Remnant Data Helps the Most?}

\begin{figure}
\centering
\includegraphics[width=\textwidth]{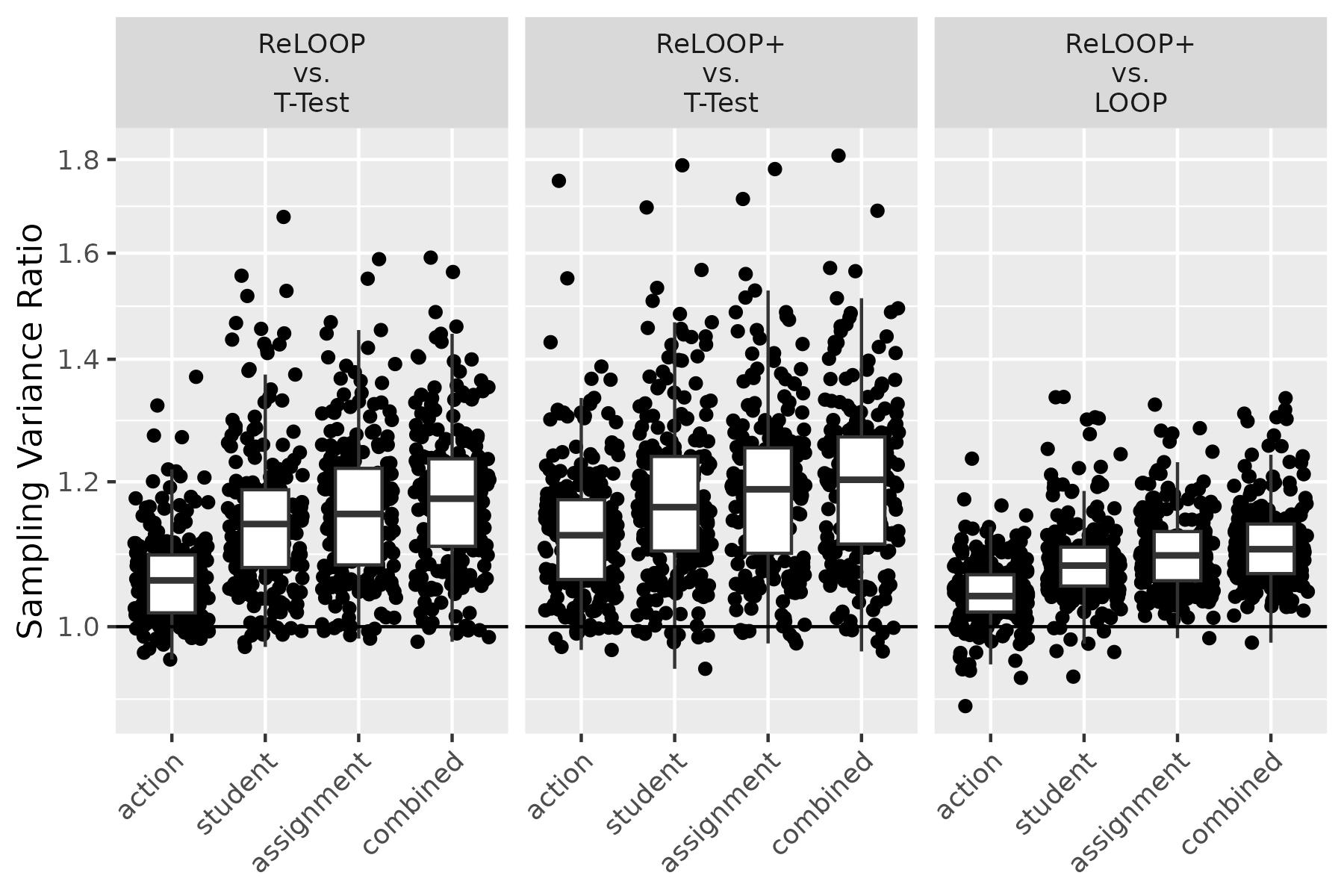}
\caption{Boxplots and jittered scatter plots of the ratios of estimated sampling variances of $\tdm$ (i.e. ``T-Test," which includes no covariate adjustment), $\myloop$ (``\textsc{loop}'', which adjusts for covariates within sample, but does not use the remnant), $\reloopV$ (``\textsc{r}e\textsc{loop},'' which adjusts for remnant-based imputations but not within-sample covariates), and $\reloop$ (``\textsc{r}e\textsc{loop}+," which adjusts for both within-sample covariates and remnant-based imputations) in 227 randomized contrasts. The Y-axis is on a logarithmic scale.}
\label{fig:restRes}
\end{figure}

Figure \ref{fig:restRes} expands on figure \ref{fig:main} by contrasting the performance of \textsc{r}e\textsc{loop} and \textsc{r}e\textsc{loop}+, relative to T-Tests and \textsc{loop}, using remnant-based imputation models trained using different types of remnant data. 
As described above, the ``action'' model uses data on each student's daily actions in ASSISTments leading up to the A/B test, the ``student'' model used student-aggregated performance metrics prior to the beginning of the A/B test, and the ``assignment'' model used student performance metrics on previous assignments or skill-builders each student had worked on. 
Finally, the ``combined'' model---also shown above,  in Figure \ref{fig:main}---was an ensemble of the action, student, and assignment models. 
By examining the performance of each separate model, we can get a sense of the relative contribution of each type of remnant data to \textsc{r}e\textsc{loop} or \textsc{r}e\textsc{loop}+'s performance. 

Comparing across models fit in the remnant, the action-level model performed the worst, while the combined model was responsible for the greatest decrease in sampling variance.
Interestingly, the assignment-level model performed nearly as well as the combined model, suggesting that action- and student-level data did not contribute substantially.
This pattern is consistent across the three different comparisons shown, comparing \textsc{r}e\textsc{loop} and \textsc{r}e\textsc{loop}+ to T-Tests, and comparing \textsc{r}e\textsc{loop}+ to \textsc{loop}. 

\section{Research Question 2: \textsc{r}e\textsc{loop} for Subgroup Effects}\label{sec:subgroup}

To judge \textsc{r}e\textsc{loop}'s potential for improving (or worsening) precision in subgroup effect estimates, we created subgroups using each of the nine student-aggregated covariates available for each of the randomized comparisons we considered. 
Specifically, we first pooled each of the 9 covariates $\bm{x}_k$, $k=1,\dots,9$ across all of the 227 pairwise comparisons, and calculated the 1/3 and 2/3 quantiles, $q_{1/3}(\bm{x}_k)$ and $q_{2/3}(\bm{x}_k)$.  
Then, for each contrast and each covariate $x$, we identified students with $x$ values that were ``low'' ($x_{ik}<q_{1/3}(\bm{x}_k)$) or ``high" ($x_{ik}>q_{2/3}(\bm{x}_k)$).
Finally, using each of the four estimators described in the previous section, for each pairwise contrast and for each covariate, we estimated two effects: one for low students and one for high. 

In addition to the nine within-sample covariates, we also looked for effects in subgroups defined by the remnant-based imputations themselves---that is, students with a high or low probability of completing their assignment, using the remnant-based model. 

All told, this should have resulted in $227\times 10 \times 2=4,540$ estimates for each of the four estimators. In practice, we did not estimate effects if either treatment arm within a subgroup had fewer than 10 subjects, which excluded 210 of these comparisons, and we encountered other estimation problems (such as the lack of variance in outcomes) in 12 others, leaving a total of 4,318 random comparisons to consider. 
Now, these 4,318 comparisons are by no means independent---they represent different ways to slice the data from the original 68 A/B tests. 
Nevertheless, by considering them all we may be able to discern some patterns in \textsc{r}e\textsc{loop}'s effectiveness in improving precision. 

\begin{figure}
\centering
\includegraphics[width=0.9\textwidth]{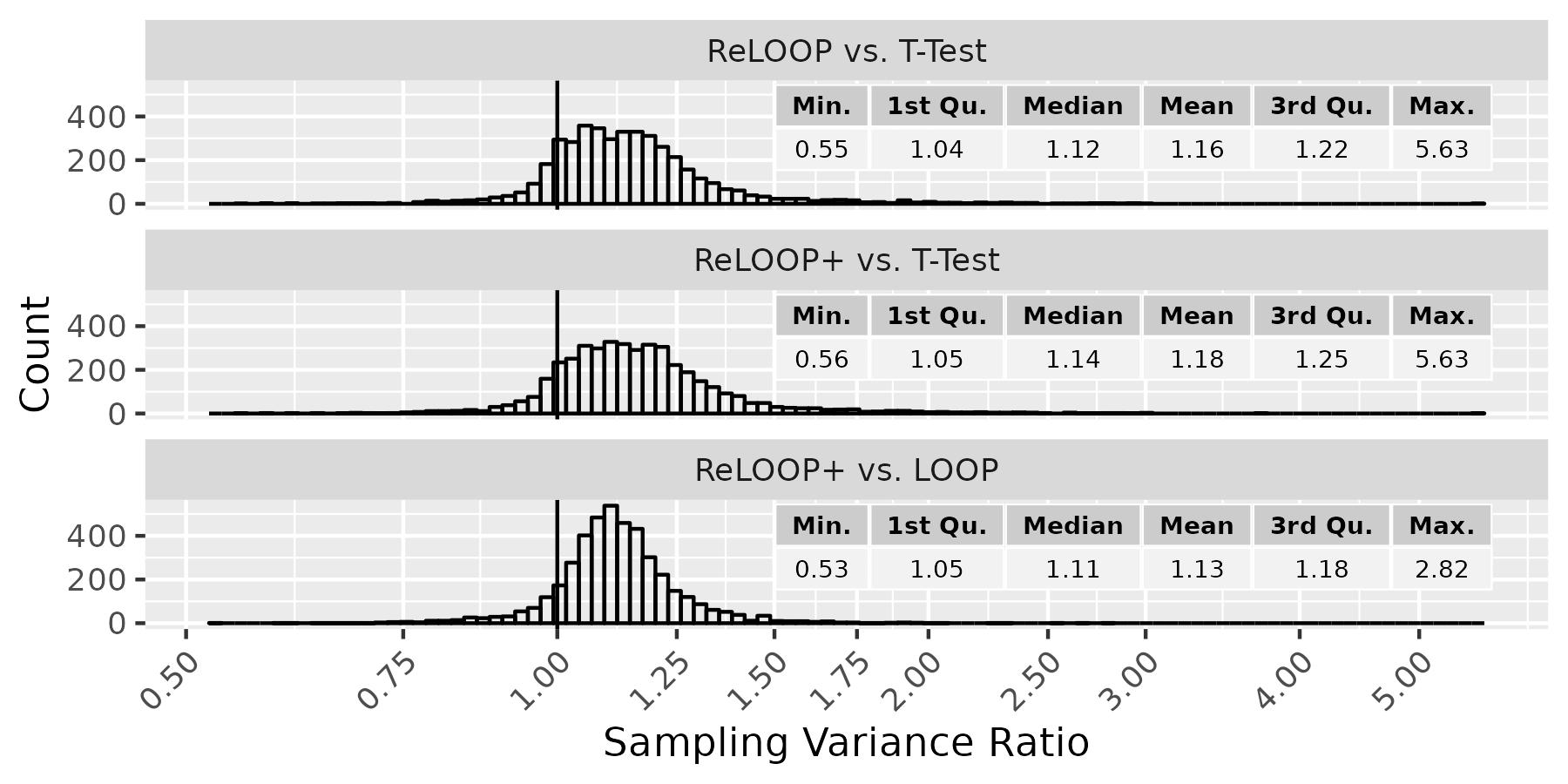}
\caption{Histograms of the ratios of sampling variances of $\tdm$ (T-Tests), $\myloop$ (\textsc{loop}), $\reloopV$ (\textsc{r}e\textsc{loop}), and $\reloop$ (\textsc{r}e\textsc{loop}+) for 4,318 estimated subgroup effects. Sample statistics of the distributions of ratios are also shown. The X-axis is on logarithmic scale.}
\label{fig:subgroupHistograms}
\end{figure}

First, though, Figure \ref{fig:subgroupHistograms} shows sampling variance ratios pooled across all A/B tests, pairwise comparisons, and subgroups. 
For the first time, we see some cases of covariate adjustment substantially harming the precision of effect estimates---\textsc{r}e\textsc{loop} gave larger standard errors than T-Tests in about 11\% of cases, \textsc{r}e\textsc{loop}+ gave larger standard errors than T-Tests in around 12\% of cases and \textsc{r}e\textsc{loop}+ gave larger standard errors than \textsc{loop} in about 13\% of cases. 
In the vast majority of these cases the effect was comparable to decreasing the sample size by less than 10\%, but in about 3\% of cases using \textsc{r}e\textsc{loop} or \textsc{r}e\textsc{loop}+ instead of T-Tests or \textsc{loop} was equivalent to decreasing the sample size by 10\% or more, and in a handful of cases the decrease was even larger, up to nearly 50\%.

Still, in the majority of cases remnant-based covariate adjustment improved the precision of impact estimates, sometimes by dramatic amounts. 
For all three comparisons shown in the figure, the median sampling variance ratio was greater than 1.1, meaning that \textsc{r}e\textsc{loop} or \textsc{r}e\textsc{loop}+ was equivalent to increasing the sample size by more than 10\% at least half the time. 
Much more dramatic improvements were also common: in 25\% of cases, \textsc{r}e\textsc{loop} outperformed the T-Test by 22\% or more, \textsc{r}e\textsc{loop}+ outperformed the T-Test by 25\% or more, and \textsc{r}e\textsc{loop}+ outperformed \textsc{loop} by at least 18\%. In some extreme cases, the improvement due to \textsc{r}e\textsc{loop} or \textsc{r}e\textsc{loop}+ was equivalent to doubling or tripling the sample size, and in one case, it was equivalent to multiplying the sample size by more than five. 

Echoing the analysis in Section \ref{sec:full-p-values}, Table \ref{tab:significanceSubgroups} shows the number of discoveries---i.e. $p<0.05$---a researcher would make using each of the three estimators. 
If p-values are not adjusted for multiple comparisons, a researcher using \textsc{r}e\textsc{loop} or \textsc{r}e\textsc{loop}+ would reject 16 or 18 more null hypotheses, respectively, than a researcher using \textsc{loop}, and 44 or 46 more than a researcher using T-Tests. 
If p-values are adjusted with the Bejamini-Hochberg procedure, a researcher using T-Tests would fail to reject every one of the 4,318 null hypotheses, while one using \textsc{loop} would reject 23, one using \textsc{r}e\textsc{loop} would reject 17, and a researcher using \textsc{r}e\textsc{loop}+ would reject 25, ensuring tenure and grant funding. 
After adjusting with the Benjamini-Yekutieli procedure, researchers using \textsc{loop} would report two discoveries, and those using \textsc{r}e\textsc{loop} or \textsc{r}e\textsc{loop}+ would report eight.

\begin{table}[ht]
\centering
\caption{The number of p-values less than $\alpha=0.05$ using each of the four estimators. The table counts significant p-values unadjusted for multiple comparisons, and adjusted with the Benjamini-Hochberg and Benjamini-Yekutieli procedures.}\vspace*{1ex}
\input{subgroupSignificance}

\label{tab:significanceSubgroups}
\end{table}

The following two subsections dig deeper into these varying effects by looking at subgroup effects broken down by subgroup and as a function of sample size. 

\subsection{Subgroup Effect Standard Errors by Covariate}

\begin{figure}
\centering
\includegraphics[width=0.9\textwidth]{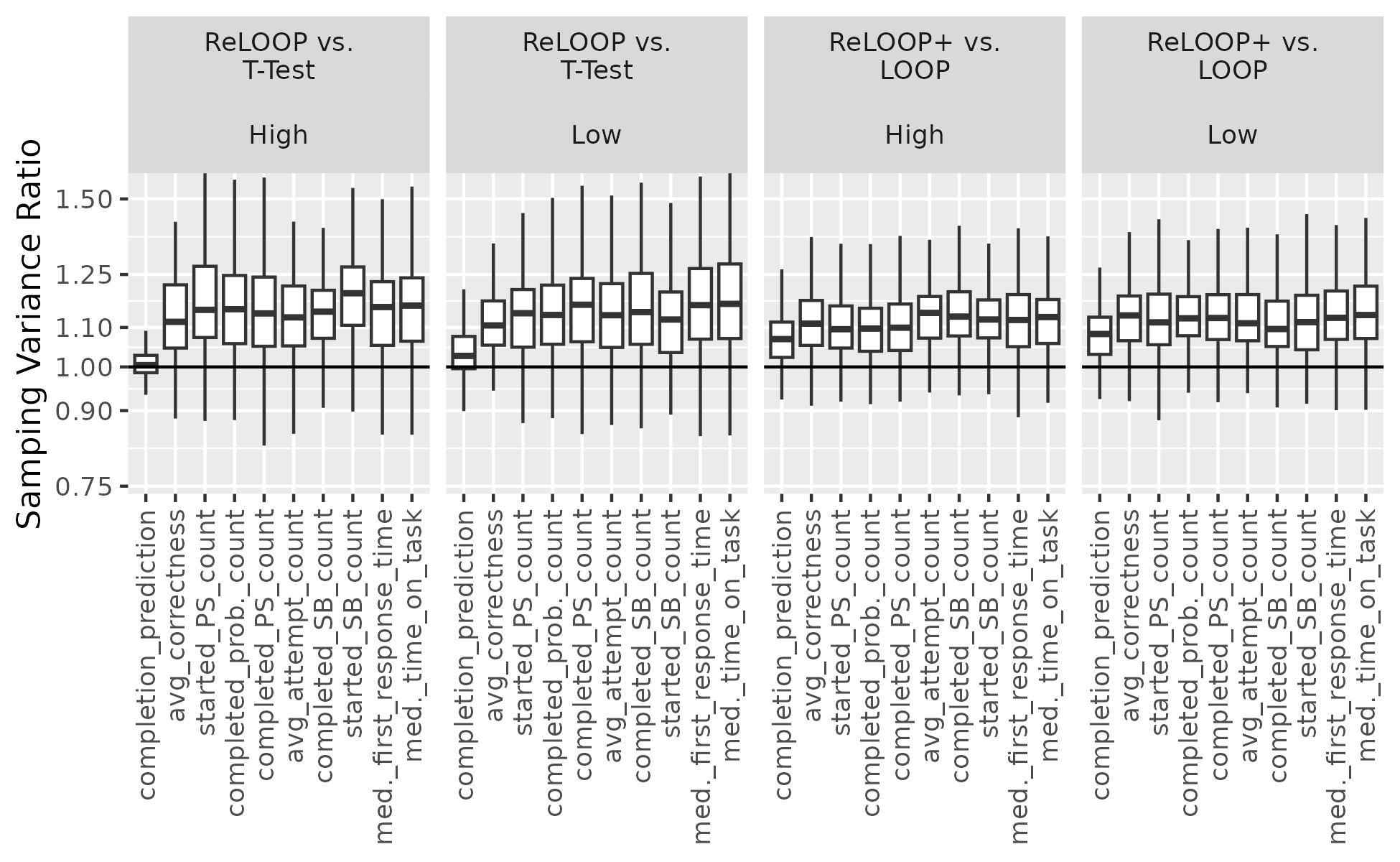}
\caption{Boxplots of the ratios of sampling variances of $\tdm$ (T-Tests), $\myloop$ (\textsc{loop}), $\reloopV$ (\textsc{r}e\textsc{loop}), and $\reloop$ (\textsc{r}e\textsc{loop}+) for each subgroup considered. Outliers are omitted. The Y-axis is on a logarithmic scale.}
\label{fig:subgroupBoxplots}
\end{figure}

Figure \ref{fig:subgroupBoxplots} shows boxplots of sampling variance ratios comparing \textsc{r}e\textsc{loop} to T-Tests and \textsc{r}e\textsc{loop}+ to \textsc{loop} for each subgroup we considered. A few features are apparent. First, \textsc{r}e\textsc{loop} performs no better than T-Tests for the high \texttt{completion\_prediction} subgroup, and little better than T-Tests for the low \texttt{completion\_prediction} subgroup. These are the subgroups defined based on $\predri$; since the variance of $\predri$ is, by definition, lower in these subgroups than in the sample as a whole, there is less opportunity to use it for variance reduction. 

Aside from those defined based on \texttt{completion\_prediction}, there was little difference in \textsc{r}e\textsc{loop}'s effectiveness between subgroups. In every case, the lower quartile was greater than 1, though the lower tail reached below 1. For comparisons between \textsc{r}e\textsc{loop} and T-Tests, the median ratio was between 1.1 and 1.2, while for \textsc{r}e\textsc{loop}+/\textsc{loop} comparisons, the medians were somewhat lower. 

\begin{figure}
\centering
\includegraphics[width=0.9\textwidth]{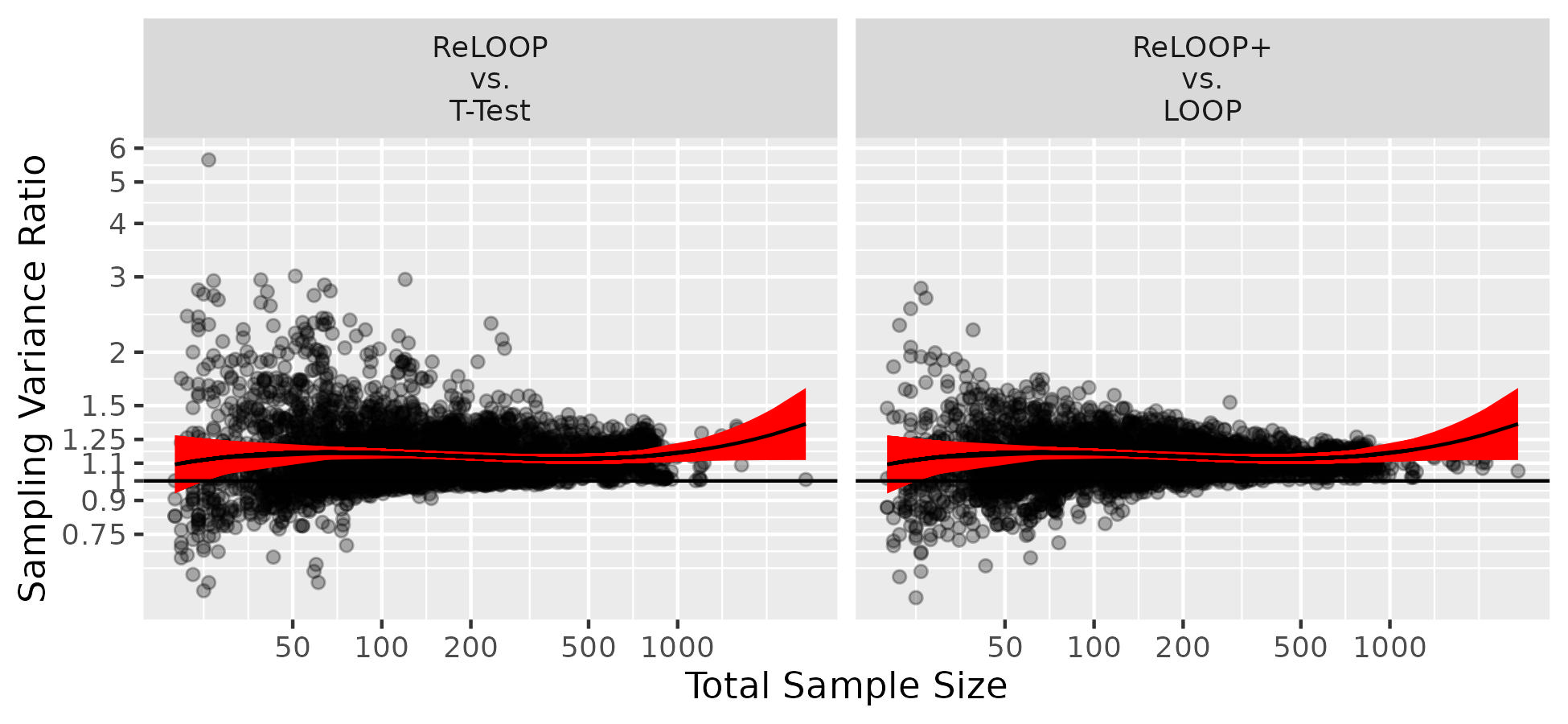}
\caption{The ratios of sampling variances of $\tdm$ (T-Tests) to $\reloopV$ (\textsc{r}e\textsc{loop}) and $\myloop$ (\textsc{loop}) to $\reloop$ (\textsc{r}e\textsc{loop}+) as the total sample size of the subgroup varies. The X- and Y-axes are on a logarithmic scale.}
\label{fig:subgroupSampleSize}
\end{figure}

Figure \ref{fig:subgroupSampleSize} plots the sampling variance ratios comparing \textsc{r}e\textsc{loop} to T-Tests and \textsc{r}e\textsc{loop}+ to \textsc{loop} against each subgroup's sample size. 
A semi-parametric regression fit (the natural logarithm of the sampling variance ratio regressed on a b-spline of the log of sample size with four degrees of freedom) is plotted over the points. The standard error shown is adjusted for the correlation of ratios from the same experiment. There is little evidence of a trend in the mean improvement due to \textsc{r}e\textsc{loop}---instead, it appears fairly constant as the sample size varies. 
On the other hand, the range and spread of ratios decrease markedly as sample sizes increase. Every case in which \textsc{r}e\textsc{loop} hurt the precision relative to T-Tests by more than 10\% was in a subgroup with $n<80$, as were all but one of the cases when \textsc{r}e\textsc{loop} adjustment was equivalent to multiplying the sample size by 2.5 or higher, relative to T-Tests. Apparently \textsc{r}e\textsc{loop}'s greatest potential for radically improving statistical precision occurs in relatively small samples. On the other hand, in relatively small samples the asymptotic guarantee that \textsc{r}e\textsc{loop} cannot increase estimated sampling variance apparently does not hold consistently.  

\section{Research Question 3: \textsc{r}e\textsc{loop} with an Unrepresentative Remnant}\label{sec:rq3}
Previous sections illustrated the potential for a model fit in the remnant to improve the precision of treatment effect estimates in A/B tests, without assuming that both datasets were drawn from the same population. 
However, in previous examples, it was not always entirely clear in what way the data from the remnant may or may not have been representative of RCT data.
In this section, we examine a case where the remnant is primarily composed of one demographic subgroup, while the RCT is a mix of subgroups.

In particular, we describe an experiment in which we intentionally designed the remnant to differ from the RCT, in order to investigate the impact remnant unrepresentativeness may have on \textsc{r}e\textsc{loop} or \textsc{r}e\textsc{loop}+'s ability to improve statistical precision. 

The experiment builds on the analyses of previous sections. 
However, to illustrate the effects of a remnant that is not representative of the RCT, we re-trained $\alg$ using a subset composed disproportionally (though not entirely) of white and Asian males, and examined the estimated sampling variance of the \textsc{r}e\textsc{loop}+ estimator for the entire RCT, for a similarly-composed subset, and for that subset's complement. 

\subsection{``Inferred Gender''}

To help maintain students' privacy, ASSISTments does not gather data on student demographics. However, the ASSISTments Foundation gathers (but does not publish) students' names, to facilitate classroom instruction (teachers need to know which student's assignment they are grading). For some analyses on ASSISTments data, analysts will attempt to guess a student's gender identification based on that student's name. To do so, the Python package ``gender-guesser''\footnote{https://pypi.org/project/gender-guesser/} was given each student's first name. The gender-guesser package uses a library of names and a script released by the German tech magazine, Heise, to determine which gender a name is associated with based on input from native speakers of various European and Asian languages. The script categorizes a name as being male, female, mostly male, mostly female, androgynous, or unknown if the name is not in the library. Clearly, this process is faulty and inexact. That being said, there is good reason to believe that most students who are inferred to be male or mostly male are male, and most inferred to be female or mostly female are female. 

There is also reason to believe that the ``unknown'' category has a higher proportion of non-Asian racial or ethnic minorities or immigrants than the inferred male or female categories. This claim follows from the assumption that names that are not in the library are uncommon and that uncommon names are probably most common among populations with non-European or non-Asian language traditions (including immigrants and native speakers with non-European or non-Asian cultural traditions) and African Americans since there is a long tradition of distinctive naming in the African American community \cite{cook2014distinctively}. 

It follows that while the set of students labeled ``Male'' or ``mostly male'' includes students with diverse genders, ethnicities, and linguistic traditions, it includes a disproportionate number of white and Asian males. In this way, this set of students follows an unfortunate, though common, pattern of disproportionately white male training sets for machine learning algorithms \cite{denton2020bringing}. 

To demonstrate the ability of the \textsc{r}e\textsc{loop}+ estimator to estimate internally-unbiased causal effects, even when the remnant reflects common biases in training datasets, we artificially limited the remnant to students labeled ``Male'' or ``mostly male''. Then, we estimated three sets of effects: one in which the RCT was limited in the same way as the remnant---i.e. to students labeled as male---another in which only the students who would be excluded from the remnant---those not labeled male---and the complete RCT data. 

\subsubsection{Results}
Using the predictions from the model described above, we estimated $\sate$ for each experimental contrast in the same four ways as in the previous sections: with T-Tests $\tdm$ (i.e. no covariate adjustment), with \textsc{loop} $\myloop$ (i.e. only within-RCT adjustment), with \textsc{r}e\textsc{loop} $\reloopV$, an estimator using \emph{only} predictions from the remnant for covariate adjustment, and with $\reloop$ \textsc{r}e\textsc{loop}+, which uses both aggregated student-level covariates and the predictions from the remnant. 

\begin{figure}
    \centering
    \includegraphics[width=0.95\textwidth]{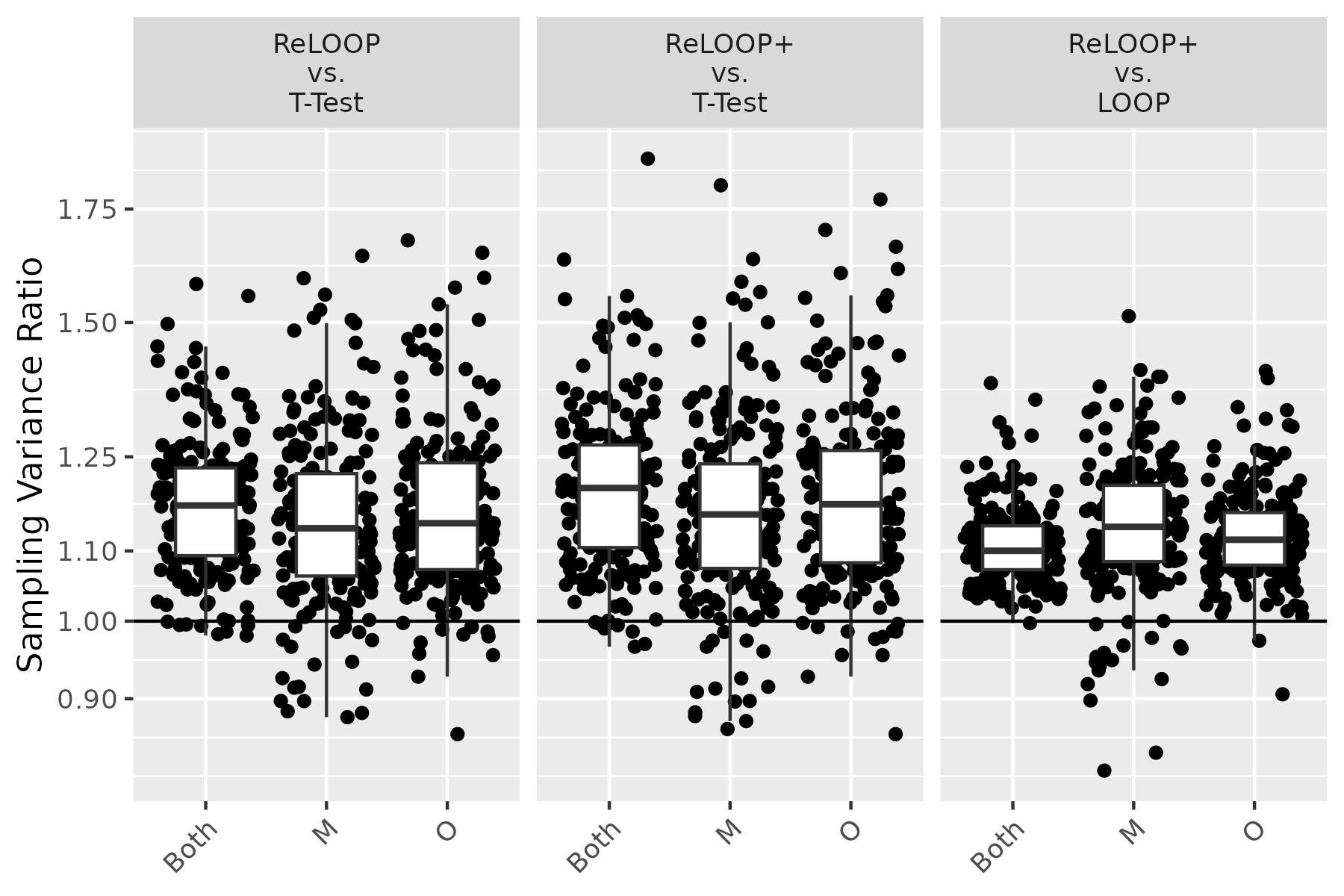}
    \caption{Results comparing estimators using imputations from the remnant, $\reloopV$ or \textsc{r}e\textsc{loop}+ (with or without other covariates), to estimators that do not, $\tdm$ and $\myloop$. For all analyses, the remnant was composed of only students whose inferred gender was male; imputations from a model trained on the male remnant were used to analyze A/B tests including all participants (``Both''), or just inferred male (``M'') or inferred non-male (``O").}
    \label{fig:results}
\end{figure}

Figure \ref{fig:results} shows the results comparing estimators that use imputations from the remnant to those that do not. 
Both estimators \textsc{r}e\textsc{loop} and \textsc{r}e\textsc{loop}+ are almost always similar to or more precise than the T-Test estimator $\tdm$.
The only exception is a handful of cases in which including remnant imputations is equivalent to decreasing the sample size by 15\% or less.
This is mostly due to very small samples in some RCTs.
On the other hand, in most of the RCTs, the improvement was 10\% or more, and in many it was upwards of 30\%.
Comparing \textsc{r}e\textsc{loop}+ to \textsc{loop}, including imputations from the biased remnant led to a 10\% or higher increase in precision in most cases.
Most surprisingly, the estimators performed as well or better in the non-Male sets and the full RCTs than in the Male subset.
That is, using a model trained in a demographically distinct population did not reduce the method's effectiveness.

\section{Research Question 4: \textsc{r}e\textsc{loop} for Population Average Effects}\label{sec:rq4}
Previous sections have focused on estimating $\sate$, the average effect of a treatment for subjects in an $\rct$. 
However, often researchers are interested in $\pate$, average effects across a wider population, $\pop$.

To attempt to estimate $\pate$, we conducted a post-stratification estimator \eqref{eq:post-stratification} using the guessed gender predictor. 
While we do not observe the true distribution of guessed gender among all middle school ASSISTments users, we may estimate it from the remnant. 
When we do so, we find that roughly a third are labeled "Male."

We calculated four post-stratified estimators for each treatment contrast, using the four sets of $\sate[G]$ estimates. 
Then, as in $\sate$ estimation, we gauged whether \textsc{r}e\textsc{loop} or \textsc{r}e\textsc{loop}+ improve the statistical precision of $\tdm$ or $\myloop$. 

\begin{figure}
    \centering
    \includegraphics[width=0.95\textwidth]{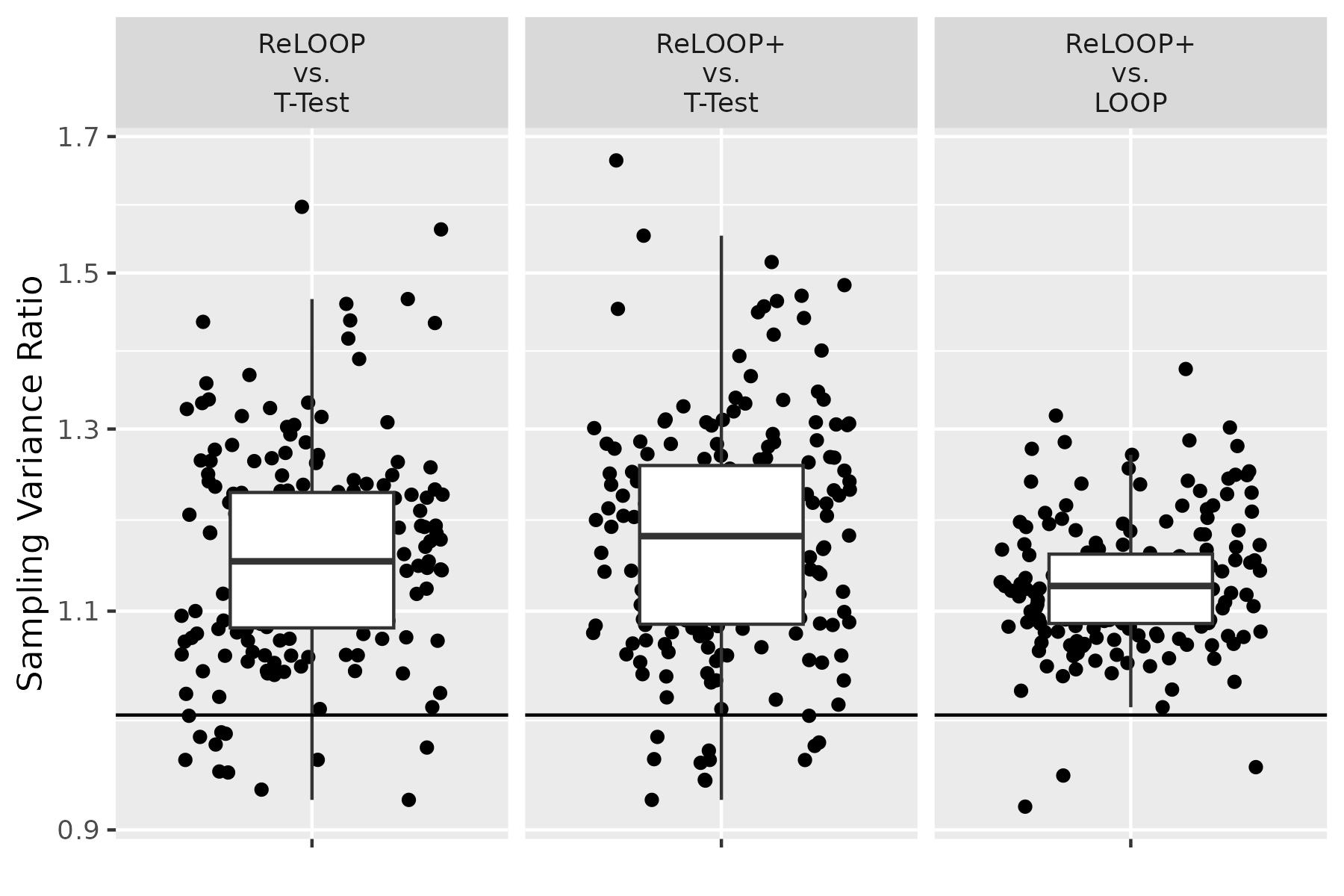}
    \caption{Results comparing post-stratification estimators using imputations from the remnants \textsc{r}e\textsc{loop} or \textsc{r}e\textsc{loop}+ (with or without other covariates) to estimators that do not, $\tdm$ and $\myloop$.}
    \label{fig:ps}
\end{figure}

Figure \ref{fig:ps} contrasts sampling variances between post-stratification using auxiliary data versus only using RCT data. 
Indeed, including imputations from the remnant improves the precision of these estimators greatly. 

While \textsc{r}e\textsc{loop} and \textsc{r}e\textsc{loop}+ are guaranteed to be internally unbiased for overall average effects $\sate$ or subgroup effects $\sate[G=k]$, these guarantees do not extend to external bias (unless, of course, $\rct$ is a random sample from $\pop$).
In particular, we cannot claim that the post-stratification estimates whose standard errors are represented in Figure \ref{fig:ps} are unbiased for $\pate$.
However, if the effects of any of the interventions in the study vary by inferred gender, then the post-stratified estimates are likely to be less externally biased than $\sate$ estimates.
We have shown here that \textsc{r}e\textsc{loop} and \textsc{r}e\textsc{loop}+ can improve their precision, as well. 

\section{Discussion}

Using remnant-trained models to predict A/B test outcomes, then using those predictions to estimate effects, has the potential to boost the precision of average effect estimators in education research. 
For typical analysis of A/B testing results, the use of remnant-based imputations could be equivalent to increasing the sample size by as much as 40-50\% relative to t-tests and as much as 30\% relative to state-of-the-art unbiased, covariate-adjusted effect estimators. 
Further, in the A/B tests we analyzed, incorporating remnant-based imputations never noticeably harmed precision. 

The benefits of remnant-based predictions were even more pronounced in estimating subgroup effects and could be roughly equivalent to increasing the sample size by factors of 2, 3, or more.
On the other hand, for subgroups with fewer than 100 students, there was a small risk that incorporating remnant-based predictions could harm precision instead of improving it. 

The benefits of using the remnant appear to extend to cases in which the remnant does not resemble data from A/B tests on demographic characteristics. In fact, counterintuitively, we found greater benefits in the subgroup that was least represented in the remnant. 

Finally, we found that incorporating remnant-based predictions into a post-stratification model can substantially improve post-stratified estimates, and hence help researchers generalize their findings to broader populations. 

\subsection{Limitations and Future Work}\label{sec:limitations}

The methods discussed here are not a panacea.
First of all, they do not apply in every randomized trial---in particular,  large datasets including covariate and outcome data for non-participants are not always available. 
Furthermore, \textsc{loop} methods are only currently available for Bernoulli or pair-randomized RCTs (including cases in which subjects are randomized with different probabilities) but not for completely randomized or cluster-randomized designs, or for general blocked designs.
We are currently working on extending \textsc{loop}---and hence \textsc{r}e\textsc{loop}---to these more complicated experimental designs, as well as to observational studies. 

Secondly, the methods may require considerable resources to implement---specifically, gathering high-dimensional covariate data from the remnant and RCT participants and formulating, tuning, and training a predictive model are all tasks that can require time, computational resources, and expertise (that said, in other work, we have seen decent precision gains from out-of-the-box random forest models). 
These issues suggest the need for guidelines as to when \textsc{r}e\textsc{loop} is likely to boost precision so that the gains in gathering and modeling remnant data will be worth the effort.

In particular, we suspect that the RCT sample size may play an important role in \textsc{r}e\textsc{loop}'s effectiveness.
Our results here suggest that the most dramatic gains from \textsc{r}e\textsc{loop} occur when the RCT sample size is below 100; however, in a handful of these cases, \textsc{r}e\textsc{loop} adjustment led to substantially higher standard errors than t-tests. 
Even when \textsc{r}e\textsc{loop} adjustment hurts precision, it does not cause bias; its associated statistical inference, such as confidence intervals and p-values, remain valid. 
Still, a method with lower chances of hurting precision, even when RCT sample sizes are small, may be desirable. 
Future research may show that simpler adjustment methods, such as \textsc{ancova}, may pose lower risks in small-sample settings. 

Prior theoretical results have shown that, regardless of the properties of an RCT, its remnant, or the imputation model, \textsc{r}e\textsc{loop} and \textsc{r}e\textsc{loop}+ cannot harm precision in large samples and that they are unbiased regardless of sample size---that is, they are unlikely to hurt an analysis. 
What this paper adds is that \textsc{r}e\textsc{loop} and \textsc{r}e\textsc{loop}+ can dramatically improve some analyses.
However, these new results are necessarily limited to the analysis of A/B tests conducted on a computer-assisted learning program, which is far from the only causal analysis in educational data mining. 
The only way to conclusively demonstrate the broad applicability and usefulness of \textsc{r}e\textsc{loop} and \textsc{r}e\textsc{loop}+ is to implement them in a wide array of contexts, perhaps alongside other causal estimators.

\section{Acknowledgements}
The research reported here was supported by the Institute of Education Sciences, U.S. Department of Education, through Grant R305D210031. The opinions expressed are those of the authors and do not represent views of the Institute or the U.S. Department of Education. 

\bibliographystyle{acmtrans}
\bibliography{bib,bib2}
\clearpage
\appendix
\section{Variables Used in Remnant Imputation Model}
\begin{table}[!h]
\caption{Prior Student Statistics Features.}
\begin{tabular}{|R{3in}|L{3in}|}
\hline
Name & Description \\
\hline
target\_sequence & The ID of the experimental skill builder \\
has\_due\_date & Whether the skill builder had a due date \\
assignments\_started & The number of assignments previously started by the student \\
assignments\_percent\_completed & The number of assignments previously completed by the student \\
median\_ln\_assignment\_time\_on\_task & The median of the log of the time between starting and finishing an assignment for all the students completed prior assignments \\
average\_problems\_per\_assignment & The average number of problems completed by the student across all their previous assignments \\
median\_ln\_problem\_time\_on\_task & The median of the log of the time the student took between starting and finished all their completed prior problems \\
median\_ln\_problem\_first\_response\_time & The median of the log of the time the student took to submit their first answer or request tutoring across all their completed prior problems \\
average\_problem\_correctness & The fraction of previously completed problems the student got correct on their first attempt without tutoring \\
average\_problem\_attempt\_count & The average number of attempts for all problems previously completed by the student \\
average\_answer\_first & The fraction of times the student submitted an answer before requesting tutoring for all problems previously completed by the student \\
average\_problem\_hint\_count & The average number of hints requested for all problems previously completed by the student \\
 
\hline
skill\_average\_problems\_per\_assignment& \multirow{7}{3in}{These features are the same as the features above with a similar name, but only calculate statistics across problems with the same skills as the problems in the experimental skill builder} \\
skill\_median\_ln\_problem\_time\_on\_task &\\
skill\_median\_ln\_problem\_first\_response\_time& \\
skill\_average\_problem\_correctness &\\
skill\_average\_problem\_attempt\_count & \\
skill\_average\_answer\_first &\\

\hline
\end{tabular}
\label{tab:pssf}
\end{table}

\begin{table}[!h]
\caption{Prior Assignment Statistics Features.}
\begin{tabular}{|R{2.5in}|L{4in}|}
\hline
Name & Description \\
\hline
id & The ID of the student \\
assignment\_start\_time & The UNIX time of when the assignment was started \\
directory\_1 & The highest level directory of the assignment location, usually an indication of curriculum \\
directory\_2 & The second level directory of the assignment location, usually an indication of grade level \\
directory\_3 & The third level directory of the assignment location, usually an indication of unit \\
sequence\_id & The unique ID of the skill builder assignment, or the corresponding normal skill builder ID for experiments \\
is\_skill\_builder & Boolean flag for whether or not this assignment is a skill builder or a normal problem set \\
has\_due\_date & Boolean flag for if the assignment has a due date \\
assignment\_completed & Boolean flag for if the student completed the assignment \\
time\_since\_last\_assignment\_start & The time between the student starting this assignment and starting their prior assignment \\
\hline
All Following Features & In addition to the raw value, a value z-scored across all students who completed the assignment previously, and a percentile across students in the same class who completed the assignment previously was included in the model as well. \\
\hline
session\_count & How many times the student left and rejoined the assignment \\
day\_count & How many days the student worked on the assignment for \\
completed\_problem\_count & How many problems the student completed in the assignment \\
median\_ln\_problem\_time\_on\_task & The median of the log of the time between the student starting and finishing problems in the assignment \\
median\_ln\_problem\_first\_response & The median of the log of the time it took for the student to submit their first answer or request tutoring on the problems they started in the assignment \\
average\_problem\_attempt\_count & The average number of attempts the student made on the problems in the assignment \\
average\_problem\_answer\_first & The fraction of times the student made an attempt before requesting tutoring on all the problems in the assignment \\
average\_problem\_correctness & The fraction of times the student got the problem correct on their first try on all the problems in the assignment \\
average\_problem\_hint\_count & The average number of hints used by the student on all the problems in the assignment \\
average\_problem\_answer\_given & The fraction of times the student was given the answer on all the problems in the assignment \\

\hline
\end{tabular}
\label{tab:pasf}
\end{table}

\begin{table}[!h]
\caption{Prior Daily Actions Features.}
\begin{tabular}{|r|l|}
\hline
Name & Description \\
\hline
id & The ID of the student \\
timestamp & The UNIX time at 00:00:00 of the day the action counts apply to \\
ln\_action\_1\_count & Log of the count of assignment started actions taken \\
ln\_action\_2\_count & Log of the count of assignment resumed actions taken \\
ln\_action\_3\_count & Log of the count of assignment finished actions taken \\
ln\_action\_4\_count & Log of the count of problem set started actions taken \\
ln\_action\_5\_count & Log of the count of problem set resumed actions taken \\
ln\_action\_6\_count & Log of the count of problem set finished actions taken \\
ln\_action\_7\_count & Log of the count of problem set mastered actions taken \\
ln\_action\_8\_count & Log of the count of problem set exhausted actions taken \\
ln\_action\_9\_count & Log of the count of problem limit exceeded actions taken \\
ln\_action\_10\_count & Log of the count of problem started actions taken \\
ln\_action\_11\_count & Log of the count of problem resumed actions taken \\
ln\_action\_12\_count & Log of the count of problem finished actions taken \\
ln\_action\_13\_count & Log of the count of tutoring set started actions taken \\
ln\_action\_15\_count & Log of the count of tutoring set finished actions taken \\
ln\_action\_16\_count & Log of the count of hint requested actions taken \\
ln\_action\_17\_count & Log of the count of scaffolding requested actions taken \\
ln\_action\_19\_count & Log of the count of explanation requested actions taken \\
ln\_action\_20a\_count & Log of the count of student correct response actions taken \\
ln\_action\_20b\_count & Log of the count of student incorrect response actions taken \\
ln\_action\_21\_count & Log of the count of open response submission actions taken \\
ln\_action\_25\_count & Log of the count of answer requested actions taken \\
ln\_action\_26\_count & Log of the count of continue selected actions taken \\
ln\_action\_30\_count & Log of the count of help requested actions taken \\
ln\_action\_31\_count & Log of the count of timer started actions taken \\
ln\_action\_32\_count & Log of the count of timer resumed actions taken \\
ln\_action\_33\_count & Log of the count of timer paused actions taken \\
ln\_action\_34\_count & Log of the count of timer finished actions taken \\
ln\_action\_35\_count & Log of the count of live tutoring requested actions taken \\
Other Actions & Artifacts of the database, always 0 \\
\hline
\end{tabular}
\label{tab:pdaf}
\end{table}



\end{document}

%% file: fullSignificance.tex
\begin{tabular}{|r|r|r|r|r|}
  \hline
 & T-Test & LOOP & ReLOOP & ReLOOP+ \\ 
  \hline
Unadjusted & 38 & 39 & 41 & 41 \\ 
  Benjamini-Hochberg & 3 & 11 & 8 & 10 \\ 
  Benjamini-Yekutieli & 2 & 2 & 2 & 2 \\ 
   \hline
\end{tabular}

%% file: subgroupSignificance.tex

\begin{tabular}{|r|r|r|r|r|}
  \hline
 & T-Test & LOOP & ReLOOP & ReLOOP+ \\ 
  \hline
Unadjusted & 375 & 403 & 419 & 421 \\ 
  Benjamini-Hochberg & 0 & 23 & 17 & 25 \\ 
  Benjamini-Yekutieli & 0 & 2 & 8 & 8 \\ 
   \hline
\end{tabular}